\documentclass[a4paperal]{i3m}

\usepackage{siunitx}
\usepackage{graphicx}
\usepackage{algorithm}
\usepackage{amsfonts}
\usepackage{amsmath}
\usepackage{wasysym}
\usepackage{algpseudocode}
\usepackage{ragged2e}

\usepackage{amsfonts}
\usepackage{amsmath}

\conference{emss}

\title{Never Mind The No-Ops: Faster and Less Volatile Simulation Modelling of Co-Evolutionary Species Interactions via Spatial Cyclic Games
}

\author{Dave Cliff}

\affil{Intelligent Systems Laboratories, School of Engineering Mathematics and Technology, University of Bristol, U.K.}

\authnote{Email address: csdtc@bristol.ac.uk}

\runningauthor{Dave Cliff}

\confyear{2024}

\begin{document}

\begin{frontmatter}
\maketitle
%
\begin{abstract}
\justifying
Issues in co-evolutionary population dynamics have long been studied via computationally intensive simulations of minimally simple agent-based models, known as Evolutionary Spatial Cyclic Games (ESCGs), involving multiple interacting biological species in which each agent has its own unique spatial location in a cell on a regular lattice, and can move from cell to cell over time. Many papers have been published exploring the dynamics of ESCGs where competitive inter-species predator/prey relationships are modelled via the cyclic game Rock-Paper-Scissors (RPS) for three species, or Rock-Paper-Scissors-Lizard-Spock (RPSLS) for five. 
At the core of these simulations is the {\em Elementary Step} (ES), in which one or two agents are chosen at random to either compete to the death, or to reproduce, or to move location. 
ESCG studies typically involve executing trillions of ESs and hence the computational efficiency of the core ES algorithm is a key concern. 
In this paper I demonstrate that the {\em de facto} standard ``Original ES'' (OES) algorithm is computationally inefficient both in time and in space due to the implicit execution of many ``no-op'' commands (i.e., commands that do nothing) and because at steady state large numbers of cells can be empty, and yet empty cells serve no purpose. 
I present a new {\em Revised ES} (RES) algorithm which eliminates these inefficiencies, and I show empirically that ESCGs with RES exhibit qualitatively the same characteristics as those with OES, and are also markedly more stable. The more stable dynamics of RES-based simulations means that they can  be run with smaller lattices than when using OES, leading to reductions in total simulation times of 85\% or more. 
Python source code developed for the experiments reported here is freely available on GitHub.     
\end{abstract}
\end{frontmatter}

\section{Introduction}
\label{sec:intro}

There is a well-established body of peer-reviewed research literature which explores issues in ecosystems stability, biodiversity, and co-evolutionary dynamics via computationally intensive simulations of minimally simple models of multiple interacting biological species. Landmark papers in this field were published in 2007--08 by \cite{reichenbach_mobilia_frey_2007_nature,reichenbach_mobilia_frey_2007_physrevlet, reichenbach_mobilia_frey_2008_jtb}, who extended the previous non-spatial model of \cite{may_leonard_1975} by modeling each species as a time-varying number of discrete individuals, where at any one time each individual occupies a particular cell in a regular rectangular lattice or grid of cells, and can {\em move} from cell to cell over time --- that is, the individuals are {\em spatially located} and {\em mobile}. Individuals can also, under the right circumstances, {\em reproduce} (asexually, cloning a fresh individual of the same species into an adjacent empty cell on the lattice); and they can also {\em compete} with individuals in neighbouring cells. Different authors use different phrasings to explain the inter-species competition, but it is common to talk in terms of predator-prey dynamics: that is, each species is predator to (i.e., {\em dominates}) some specified set of other species, and is in turn also prey to (i.e., is {\em dominated} by) some set of other species. 

\begin{figure}
\begin{center}
\includegraphics[trim=0cm 0cm 0cm 0cm,clip=true,scale=0.35]{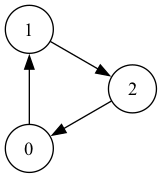}
\end{center}
\caption{Dominance network, a directed graph or {\em digraph}, for the three-species Rock-Paper-Scissors (RPS) game. Species $S_i$ are denoted by nodes numbered by index $i\in\{1, 2, 3\}$, with directed edges running from the dominator (``predator'') species to the dominated (``prey''') species. There are multiple {\em labelings} of this graph (e.g.: $(S_1$$=$$R, S_2$$=$$P, S_3$$=$$S); (S_1$$=$$P, S_2$$=$$S, S_3$$=$$R); \ldots$) but if one graph can be turned into another purely by rearranging the node labels then those two graphs are topologically equivalent or {\em isomorphic}. All possible labelings of the RPS digraph are isomorphic with each other, so there is only one isomorphically unique RPS digraph.}
\label{fig:RPSnet}
\end{figure}

The population dynamics are determined to a large extent by the model's {\em dominance network}, a directed graph (digraph) where each node in the network represents one of the species in the model, and a directed edge (i.e., an arrow) from the node for species $S_i$ to the node for species $S_{j: j\neq i}$ denotes that $S_i$ dominates $S_j$. To exhibit interesting long-term dynamics, the dominance network must contain at least one cycle (i.e., a path from some species $S_i$ to some $S_j$ traced by traversing edges in the directions of the arrows, potentially passing through some number of intermediate species' nodes, where at $S_j$ there then exists an as-yet-untraversed edge back to $S_i$). Under the constraint that no species can be both predator and prey to another species at the same time, the smallest dominance network of interest is a minimal three-node cycle, which represents the intransitive dominance hierarchy of the simple hand-gesture game Rock-Paper-Scissors (RPS), as illustrated in Figure~\ref{fig:RPSnet}. In RPS-based models, the three species are $R$ (rock), $P$ (paper), and $S$ (scissors) and when two neighboring individuals compete the rules are as follows: if they are both the same species, the competition is a draw and nothing else happens; but otherwise $R$ kills $S$, $S$ kills $P$, and $P$ kills $R$, with the cell where the killed individual was located being set to empty, denoted by $\emptyset$. Because the individuals in these models are spatially located on a lattice, and because the inter-species competition is determined by having pairs of individuals play RPS-like games with cyclic dominance digraphs, this class of co-evolutionary population dynamics models is often referred to as  {\em evolutionary spatial cyclic games} (ESCGs). 

ESCGs are inherently stochastic and to generate rigorous results it is often necessary to simulate ESCG systems many times, aggregating results over many independent and identically distributed (IID) repetitions of the system evolving over time.  At the core of these simulations is the {\em Elementary Step} (ES), in which one or two agents are chosen at random to either compete to the death, or to reproduce, or to move location. 
ESCG studies typically involve executing trillions of ESs and hence the computational efficiency of the core ES algorithm is a key concern. 
In this paper I demonstrate that the {\em de facto} standard ``Original ES'' (OES) algorithm is computationally inefficient both in time and in space due to the implicit execution of many ``no-op'' commands (i.e., commands that do nothing) and because at steady state large numbers of cells can be empty, and yet empty cells serve no purpose. 

The rest of this paper is structured as follows. Section~\ref{sec:ESCGs} gives more details of the background, specifically of the original elementary step as it is usually described in the literature. Section~\ref{sec:probabilities} discusses the interpretation of three key parameters in ESCGs. Section~\ref{sec:algorithms} first explicitly re-states the OES as an algorithm, and highlights the multiple inefficiencies that lurk within it, and then introduces my {\em Revised Elementary Step} (RES). In Section~\ref{sec:results}, I present empirical results which indicate that the switch to RES offers definite benefits, chief of which is a reduction in volatility of population dynamics, which means that useful results can be generated from RES working with smaller lattices, giving major savings in overall computation time. The results are discussed in Section~\ref{sec:discussion} and conclusions drawn in Section~\ref{sec:conclusion}.
 
 \section{Evolutionary Spatial Cyclic Games (ESCGs)}
 \label{sec:ESCGs}
 
Almost all simulations of co-evolutionary population dynamics via ESCGs are simple discrete-time systems that are technically unchallenging to write a program for, and are strongly reminiscent of -- but not identical to -- cellular automata (see e.g. \cite{wolfram_2002_book}). The lattice/grid needs to first be set up, i.e.\ its dimensions and initial conditions at the first time-step need to be specified. Each cell in the grid is either empty, or contains exactly one individual organism, and each individual is a member of exactly one of the model's set of species. If the number of species in the model is denoted by $N_s$, one common style of initialisation is to assign one individual to every cell in the grid, with that individual's species being an equiprobable choice from the set of available species (i.e., choose species $S_i$ with probability $1/N_s; \forall i$). The modeller also needs to specify the dimensionality of the lattice, and its {\em length} (i.e., number of cells) along each dimension. In almost all of the published work  in this field, the lattice is two-dimensional and square, so its extent is defined by a single system hyperparameter: the side-length (conventionally denoted by $L$). The total number of cells in the lattice (conventionally denoted by $N$) is hence $N=L^{2}$. Working with 2D lattices has the advantage that the global state of the system can be readily visualised as a snapshot at time $t$ as a color-coded or gray-shaded 2D image, with each species in the model assigned its own specific color or gray-scale value, and animations can easily be produced visualising the change in the system state over time. 

In the literature on ESCGs, authors often make the distinction between two scales of time-step in the simulation. At the very core of the simulation process is a loop that iterates over a number of {\em elementary steps} (ESs), the finest grain of time-step; and then some large number of consecutive ESs is counted as what is conventionally referred to as a {\em Monte Carlo Step} (MCS).  

In a single ES, one individual cell (denoted $c_i$) is chosen at random, and then one of its immediately neighboring cells (denoted $c_n$) is also chosen at random: in almost all of the literature on 2D lattice model ESCGs, the set of neighbours is defined as the 4-connected von Neumann neighborhood rather than the 8-connected Moore neighborhood commonly used in cellular automata research (although e.g.\ \cite{laird_schamp_2006_RPSLS,laird_schamp_2008_RPSLS,laird_schamp_2009_coexistence_RPSLS} used the Moore neighborhood in their lattice models), and the work reported here uses von Nuemann. There seems to be no firm convention on whether to use periodic boundary conditions (also known as toroidal wrap-around) or ``walled garden'' no-flux boundary conditions (such that cells at the edges and corners of the lattice have a correspondingly reduced neighbour-count) -- some authors use periodic, others no-flux. The results presented in this paper come from simulations with no-flux boundary conditions. 

In each ES one of three possible actions occurs: {\em competition}, {\em reproduction}, or {\em movement}, and the probabilities of each of these three actions occurring per ES is set by system parameters $\mu, \sigma,$ and $\epsilon$, respectively (this is explained in more precise detail later, in Section~\ref{sec:probabilities}). {\em Competition} involves the individuals at $c_i$ and $c_n$ interacting according to the rules of the cyclic game, resulting in either a draw or one of the individuals losing, in which case it is deleted from its cell, replaced by  $\emptyset$; {\em reproduction} occurs when one of $c_i$ or $c_n$ holds $\emptyset$, the empty cell being filled by a new individual of the same species as the nonempty neighbor; and {\em movement} involves swapping the contents of $c_i$ and $c_n$.  

Because, in the original formulation, each ES involves only one of the three possible actions (competition, reproduction, or movement) occurring for a single cell, a Monte Carlo Step (MCS) is conventionally defined as a sequence of $N$ consecutive ESs, the rationale being that, on the average, each cell in the lattice will be randomly chosen once per MCS, and hence that, again on the average, every cell in the grid has the potential to change once between any two successive MCSs. Most published research on this type of model uses MCS as the unit of time when plotting time-series graphs illustrating the temporal evolution of the system, and I follow that convention here. Some authors don't refer to MCS but instead talk of each sequence of $N$ consecutive ESs in their ESCG as one new {\em generation}. 

In their seminal papers, \cite{reichenbach_mobilia_frey_2007_nature,reichenbach_mobilia_frey_2007_physrevlet,reichenbach_mobilia_frey_2008_jtb} studied 2D lattice systems where interspecies competition was via $N_s$$=$$3$ RPS games, with $\mu$$=$$\sigma$$=$$1.0$, and where $L$ ranged from 100  to 500, and they showed and explained how the overall system dynamics result in emergence of one or more temporally and spatially coherent interlocked {\em spiral waves}.
The specific nature of the wave-patterning, i.e. the size and number of spiral waves seen in the system-snapshot images, depended on a {\em mobility} measure $M$$=$$\epsilon/2N$, which is proportional to the expected area of lattice explored by a single agent  per MCS. Given that $\epsilon$ is a probability and hence $\in[0.0,1.0]$, the largest meaningful $M$ value is  $M_{\text{max}(L)}=1/2L^2$.

In the years since publication of \cite{reichenbach_mobilia_frey_2007_nature,reichenbach_mobilia_frey_2007_physrevlet, reichenbach_mobilia_frey_2008_jtb}, many papers have been published that explore the dynamics of such co-evolutionary spatial RPS models. For examples of recent publications exploring a range of issues in the three-species RPS ESCG, see: \cite{
nagatani_ichinose_tainaka_2018_RPS,
kabir_tanimoto_2021_RPS,
mohd_park_2021_RPS,
bazeia_bongestab_deoliveira_2022_RPS,
park_2021_RPS,
menezes_batista_rangel_2022_RPS,
menezes_rangel_moura_2022_RPS,
zhang_bearup_guo_zhang_liao_2022_RPS,
menezes_barbalho_2023_RPS,park_jang_2023_RPS}; and \cite{kubyana_landi_hui_2024_RPS}.

\begin{figure}
\begin{center}
\includegraphics[trim=0cm 0cm 0cm 0cm, clip=true,scale=0.35]{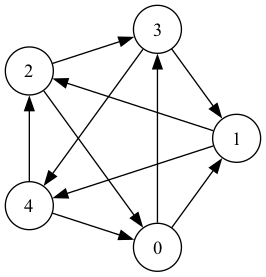}
\end{center}
\caption{Dominance network digraph for the five-species Rock-Paper-Scissors-Lizard-Spock (RPSLS) cyclic game. The rules of this game are: scissors cut paper;
paper covers rock; 
rock blunts scissors;
scissors decapitates lizard;
lizard eats paper;
paper disproves Spock;
Spock vaporizes rock;
rock crushes lizard;
lizard poisons Spock; and
Spock smashes scissors.
}
\label{fig:RPSLSnet}
\end{figure}

More recently, various authors have reported experiments with a closely related system where $N_s$=$5$: this game is known as Rock-Paper-Scissors-Lizard-Spock (RPSLS), an extension of RPS introduced by \cite{kass_bryla_1998}  and subsequently featured in a 2012 episode of the popular US TV show {\em Big Bang Theory}. The dominance network for the RPSLS game is illustrated in Figure~\ref{fig:RPSLSnet} and explained in the caption to that figure. This (and other five-species ESCGs) was first explored in the theoretical biology literature by \cite{laird_schamp_2006_RPSLS,laird_schamp_2008_RPSLS,laird_schamp_2009_coexistence_RPSLS}; and RPS-like ESCGs with $N_S$$\geq$$5$ were explored by \cite{avelino_deoliveria_trintin_2022_RPS_bigN}.

In a recent paper, \cite{zhong_etal_2022_ablatedRPSLS} studied the effects on co-evolutionary dynamics of selectively ablating the RPSLS dominance network, i.e., deleting one or more of the directed edges in the RPSLS digraph and exploring the consequent changes in the population dynamics. In the abstract to their paper, Zhong et al.\ wrote that these systematic changes to the dominance network (which they refer to as the {\em interaction structure}):
\begin{quotation}
 ``$\ldots$ impacts the evolutionary dynamics, and different interaction structures allow for different states of multi-species coexistence. We also find that the competition between different three-species-cyclic interactions is crucial for the realization of different asymptotic behaviors at low mobility. Our findings may be useful to understand the subtle effects of competitive structure on species coexistence and evolutionary game outcomes."  
\end{quotation}

Here I will show that the simulation algorithm used by Zhong et al., and seemingly many other authors whose work I have cited in this paper, is inefficient and can be revised to give reductions in total computation time of 80\% or more, with the results still showing the same qualitative features as identified in \cite{zhong_etal_2022_ablatedRPSLS}. I also argue that Zhong et al.'s failure to report any preparatory baseline tests on unablated dominance networks casts some doubt on their results. 

Let $N_a$ denote the number of ablated edges in the dominance network. Zhong et al.\ show (in their Figures~3, 5, 6, and~7) results from many thousands of independent and identically distributed (IID) repetitions of simulations of the ablated-digraph RPSLS systems for networks with $N_a$$\in$$\{1, 2, 3\}$ ablated edges, plotting the frequencies of different classes of outcome. Zhong et al.\ run their simulations for 100,000MCS (which hereinafter I'll write as 100kMCS) and refer to the outcome of their experiments, the state of the system after 100kMCS, as the  ``asymptotic state'', although later in this paper I present evidence that their system can be far from an asymptote, far from a steady state, after 100kMCS. 

The treatment that Zhong et al.\ vary in their experiments is the mobility parameter $M$, which they vary over the range $M$$\in$$[10^{-7}, 10^{-3}]$ in their Figures~3, 5, 6, and~7, and the response that they monitor is the number of species remaining at the end of the experiment. Here, let  $n_s(t)$ denote the number of species remaining at time $t$ (where the unit of time is MCS), so Zhong et al.'s key metric is $n_s(100\text{k})$. Zhong et al.'s  primary observation is that for each value of $N_a$,  at low $M$, the system almost always converges to $n_s(100\text{k})$$=$$3$, and then as $M$ is increased the system shows a sudden and steep decline in frequency of $n_s(100\text{k})$$=$$3$ and instead $n_s(100\text{k})$$\leq$$2$ is the outcome in $\approx$$100\%$ of the experiments. In all of Zhong et al.'s  experiments, this sudden transition, the change in $n_s(100\text{k})$  from three to two or fewer, occurs very sharply in the interval $M$$\in$$[10^{-5},10^{-4}]$. 

Zhong et al.\ used a square grid with L=200. For each data-point on their asymptotic-state frequency plots they computed 500 IID repetitions of any one experiment for any given value of $M$, and each plot has data-point markers showing that they sampled 20 different values of $M$. Thus, in total, the four asymptotic-state frequency figures presented by Zhong et al.\ represent results from $4 \times 20 \times 500 \times 10^{5} = 4.0$$\times$$10^9$ individual MCS, and each of their MCS involves $L^2$$=$$4$$\times$$10^5$ ESs, so the total number of ESs they simulated is $1.6$$\times$$10^{14}$ -- that's 160 trillion elementary steps.

My work reported in this paper grew from an attempt to replicate and extend the results of Zhong et al.: in doing that, I wrote my own simulation of their evolutionary spatial cyclic game (ESCG) system in Python. And in doing that, I came to realise that the original specification of these systems given verbally by \cite{reichenbach_mobilia_frey_2007_nature,reichenbach_mobilia_frey_2007_physrevlet, reichenbach_mobilia_frey_2008_jtb}, (i.e., written in English as a series of sentences, rather than being formally expressed as an algorithm in pseudocode or a known programming language), which was then faithfully copied by many other authors including  \cite{zhong_etal_2022_ablatedRPSLS}, was somewhat ambiguous and was implicitly significantly computationally inefficient in space and in time. Once expressed as an algorithm rather than as English text, it became easy to see that a more efficient ES algorithm was likely to give very similar results in much less elapsed computation time than the original. This paper is devoted to explaining my {\em Revised ES} (RES) algorithm, and to demonstrating that it does indeed offer significant savings in compute-time.

This paper concentrates on the algorithmic and implementational aspects of RES, and on demonstrating sufficient similarity between results from the RES-based simulation, and my implementation of the {\em Original ES} (OES)-based system used by Zhong et al. and many others. In a companion paper \cite{cliff_2024_circulants} I show the RES-based simulator being used to explore a series of major extensions to the work of Zhong et al., but this current paper focuses only on replicating in RES those OES simulation results previously published by \cite{zhong_etal_2022_ablatedRPSLS}.

\section{Discussion: what exactly are $\mu$, $\sigma$, and $\epsilon$?}
\label{sec:probabilities}

Many of the background details necessary to understand the work reported here have already been discussed in the narrative introduction of the previous section. However, there is one point touched upon briefly in Section~\ref{sec:intro} that warrants further discussion, and that concerns the interpretation of the system parameters  $\mu$, $\sigma$, and $\epsilon$. 

\cite{zhong_etal_2022_ablatedRPSLS} state in Section~2 of their paper:

\begin{quote}
``In each elementary time step, an individual on one node [i.e., in one cell] and one of its neighboring nodes [cells] are randomly selected. Three possible actions [i.e., competition, reproduction, and migration] \ldots occur between the two selected nodes [cells] at the probabilities $\mu$, $\sigma$, and $\epsilon$, respectively.  \ldots In each elementary time step, only one action occurs. One full Monte Carlo step (MCS) contains $N$ elementary time steps \ldots in order to guarantee each individual could be selected for interaction once on average. \ldots Without loss of generality, we set $\mu$$=$$\sigma$$=$$1$, and $\epsilon$$=$$2$$\times$$M$$\times$$N$.'' 
\end{quote}

\noindent
And their paper then says no more about $\mu$, $\sigma$, or $\epsilon$. 

Their phrasing here is unfortunate, and obscures an important detail. If $\mu$, $\sigma$, and $\epsilon$ really are probabilities, then presumably we could set each to 0.5 and intuitively assume that on any one elementary step (ES) there is a 50\% chance of each of the three actions occurring --- but the expected number of actions per ES would then be 1.5, and one eighth of all ESs would have all three actions occurring in them, and both those outcomes contradict Zhong et al.'s assertion that in each ES, only one action occurs. 

Other authors working on this class of ESCG 2D lattice model have been more forthcoming. In describing their ESCG using RPSLS, \cite{cheng_yao_huang_park_do_lai_2014} write that they ``\ldots normalize the total probability of three actions into unity.'', but they give no specific details of this normalization step. 

The normalization step is clearly explained by \cite{park_jang_2019} who studied a 2D ESCG RPSLS system which involved an additional probability $\gamma$ for the rate of one half of the space of possible cell-to-cell competitions, the other half's rate remaining set by  $\sigma$.  As is conventional in this field, \cite{park_jang_2019} use the symbol $\Box$ as a `wildcard', matching to any cell in the lattice regardless of its contents, and explain their system via four relations as follows:

\begin{quote}

\dots
In the spatial dynamics of RPSLS game, five species (referred to as A, B, C, D, and E) randomly populate on
a square lattice with periodic boundary conditions and interact each other in the nearest neighborhood according to the set of following
rules:
\begin{equation}
AB \xrightarrow{\sigma} A\emptyset,
BC \xrightarrow{\sigma} B\emptyset,
CD \xrightarrow{\sigma} C\emptyset,
DE \xrightarrow{\sigma} D\emptyset,
EA \xrightarrow{\sigma} E\emptyset,
\end{equation}
\vspace*{-8mm}
\begin{equation}
AD \xrightarrow{\gamma} A\emptyset,
BE \xrightarrow{\gamma} B\emptyset,
CA \xrightarrow{\gamma} C\emptyset,
DB \xrightarrow{\gamma} D\emptyset,
EC \xrightarrow{\gamma} E\emptyset,
\end{equation}
\vspace*{-8mm}
\begin{equation}
A\emptyset \xrightarrow{\mu} AA,
B\emptyset \xrightarrow{\mu} BB,
C\emptyset \xrightarrow{\mu} CC,
D\emptyset \xrightarrow{\mu} DD,
E\emptyset\xrightarrow{\mu} EE,
\end{equation}
\vspace*{-8mm}
\begin{equation}
A\Box \xrightarrow{\epsilon} \Box A,
B\Box \xrightarrow{\epsilon} \Box B,
C\Box \xrightarrow{\epsilon} \Box C,
D\Box \xrightarrow{\epsilon} \Box D,
E\Box \xrightarrow{\epsilon} \Box E,
\end{equation}

\ldots
All relations (1)–(4) can actually occur only when the states of both sites [cells] meet the requirement for the particular interaction
with normalized probabilities:
$
 \sigma / (\sigma +\gamma +\mu + \epsilon), 
 \gamma / (\sigma +\gamma +\mu + \epsilon), 
 \mu/ (\sigma +\gamma +\mu + \epsilon),$ and
$ \epsilon/ (\sigma +\gamma +\mu + \epsilon)$, respectively\ldots
\end{quote}

Zhong et al.\ explain their model using very similar notation to Park \& Jang, but Zhong et al.\ use the single parameter $\sigma$ for the rate of all competitions, so Zhong et al.'s Relation 2 has $\sigma$ as its parameter, instead of $\gamma$.

From this it seems reasonable to infer that in \cite{zhong_etal_2022_ablatedRPSLS}, the actual operational probability of competition (denoted here by ${\text{Pr}}({\cal C})$) occurring in any one ES is given by  $\sigma / (\sigma$$+$$\mu$$+$$\epsilon)$; the probability of reproduction by $\text{Pr}({\cal R}) = \mu / (\sigma$$+$$\mu$$+$$\epsilon)$; and the probability of movement by $\text{Pr}({\cal M}) = \epsilon / (\sigma$$+$$\mu$$+$$\epsilon)$. All my experiments reported in Section~\ref{sec:results} are generated from an implementation of this interpretation. 

To give an illustrative example: later, in Section~\ref{sec:results}, I show results from simulations using OES where $\mu$$=$$\sigma$$=1$,  $L$$=$$200$, and where $M$ ranges up to $M$$=$$10^{-3}$, the largest value explored in \cite{zhong_etal_2022_ablatedRPSLS}.  Now $M$$=$$10^{-3}$ is manifestly larger than $M_{\text{max}(200)}$$=$$1.25$$\times$$10^{-5}$ ($M_{\text{max}(L)}$ was introduced in Section~\ref{sec:ESCGs})
and hence, {\em prima facie}, this results in an out-of-range value for $\epsilon$,  
given that $\epsilon$ is defined as a probability 
(i.e., here 
$\epsilon=2MN= 2$$\times$$(1$$\times$$10^{-3}$$)\times$$(4.0$$\times$$10^{5}) = 8$$\times$$10^{2} \notin [0.0,1.0]$). 
However, after the ``normalisation'' process, the actual OES probabilities are $\text{Pr}({\cal C})$$=$$\text{Pr}({\cal R})$$=$$1/(2+\epsilon);$  and $\text{Pr}({\cal M})$$=$$\epsilon/(2+\epsilon)$, which in this case works out at $\text{Pr}({\cal C})$$=$$\text{Pr}({\cal R})$$=$$0.00125$ and $\text{Pr}({\cal M})$$=$$0.9975$. 
When using RES, the probabilities are simply set explicitly, i.e.\ the corresponding RES experiments would use $L$$=$$200$, $\mu$$=$$\sigma$$=$$0.00125$, and $\epsilon$$=$$0.9975$. For this reason, later in this paper when results are presented to compare OES outcomes with RES outcomes, the $M$-values along the horizontal axis on the OES graphs run up to values higher than $M_{\text{max}(L)}$, whereas in the corresponding RES graphs there is a flatline in response past $M$$>$$M_{\text{max}(L)}$ because $\epsilon$$=$$1.0$ for any $M$$>$$M_{\text{max}(L)}$.

\section{Replicating \& Revising the Elementary Step}
\label{sec:algorithms}

\subsection{Reimplementation of Zhong et al.'s ESCG}

Let $l$ be the 2D lattice of cells such that an individual cell with coordinates $(x,y)$ in the lattice is denoted by $l(x,y)$, and where the individual cell is the {\em position} of an individual agent $i$ in the model, I'll denote that by $\vec{p}_i$. 
Let  ${\cal U}$$\left[n_{lo},n_{hi}\right]$ denote a new random draw from a uniform distribution over the range $\left[n_{lo},n_{hi}\right]$$\in$${\cal R}$ and similarly let ${\cal U}\{ m_0, m_1, \ldots\}$ represent a new uniform (equiprobable) random choice of member from the set $\{ m_0, m_1, \ldots\}$.

Assume here the existence of a function {\sc RndNeighbour}$(l, \vec{p}_i, {\cal N}, {\cal B})$ which returns the coordinate pair $\vec{p}_n=(x_n, y_n)$ for a randomly chosen member from the neighbourhood of $\vec{p}_i=(x_i,y_i)$ where ${\cal N}$ specifies the neighborhood function to use  (e.g. von Neumann or Moore, etc), and with boundary conditions specified by ${\cal B}$. 

Assume also that the four RPSLS relations introduced above in the quote from \cite{park_jang_2019}, denoted here by $R1$ to $R4$,  are encoded as three functions, each of which take as arguments the lattice $l$, the lattice position $\vec{p}_i$ of the randomly chosen individual cell $c_i$, and the lattice coordinates $\vec{p}_n$ of $c_i$'s randomly chosen neighboring cell, denoted $c_n$:
\begin{itemize}

\item {\sc Compete}$(l,\vec{p}_i,\vec{p}_n)$ executes $R1$ and $R2$. 

\item {\sc Reproduce}$(l,\vec{p}_i,\vec{p}_n)$ executes $R3$. 

\item {\sc Move}$(l,\vec{p}_i,\vec{p}_n)$ executes $R4$. 

\end{itemize}

\noindent
Each of these functions returns the updated lattice $l$. 
These three functions will be called from within the procedure for a single elementary step, {\sc ElStep} which takes as arguments the lattice, the position vectors $\vec{p}_i$ and $\vec{p}_n$ of $c_i$ and $c_n$ respectively, and the three control probabilities $\mu, \sigma,$ and $\epsilon$, and returns the updated lattice. The Elementary Step (ES) is listed in Algorithm~\ref{alg:OES}, and then the entire algorithm for an instance of the evolutionary spatial cyclic game (ESCG) on a square 2D lattice is as shown in Algorithm~\ref{alg:Game}.

\begin{algorithm}
\caption{Original Elementary Step (OES)}
\label{alg:OES}
\begin{algorithmic}[1]
\Require $\mu \in [0.0, 1.0] \subset {\mathbb R}$ \Comment{Pr(compete)}
\Require $\sigma \in [0.0, 1.0] \subset {\mathbb R}$ \Comment{Pr(reproduce)}
\Require $\epsilon \in [0.0, 1.0] \subset {\mathbb R}$ \Comment{Pr(move)}
\Require $l$ \Comment{Current state of lattice $l$}
\Require $\vec{p}_i$ \Comment{Lattice coords of cell $c_i$}
\Require $\vec{p}_n$ \Comment{Lattice coords of cell $c_n$}
\Procedure{ElStep}{$l, \vec{p}_i, \vec{p}_n, \mu, \sigma, \epsilon$} 

	\State$\bar{\mu} \gets {\mu}/(\mu + \sigma + \epsilon)$\Comment{``normalize'' $\mu$}
	\State$\bar{\sigma} \gets {\sigma}/(\mu + \sigma + \epsilon)$\Comment{``normalize'' $\sigma$}	
	\State$a \gets {\cal U}\left[0.0,1.0\right]$
	\If{$0 \leq a < \bar{\mu} $}
    			\State $l \gets ${\sc Compete}$(l,\vec{p}_i,\vec{p}_n)$
    	\ElsIf{$\bar{\mu} \leq a < (\bar{\mu} + \bar{\sigma})$}
		    \State $l \gets ${\sc Reproduce}$(l,\vec{p}_i,\vec{p}_n)$
	\Else
    			\State $l \gets ${\sc Move}$(l,\vec{p}_i,\vec{p}_n)$
	\EndIf
	\State \textbf{return} {$l$}
\EndProcedure
\end{algorithmic}
\end{algorithm}

\begin{algorithm}[t]
\caption{Evolutionary Spatial Cyclic Game (2D Square)}
\label{alg:Game}
\begin{algorithmic}
\Require $L \geq 1 \in {\mathbb Z}$ \Comment{Side-length of square lattice $l$}
\Require $M \in (0.0, 1.0) \subset {\mathbb R}$ \Comment{Mobility}
\Require $\mu \in [0.0, 1.0] \subset {\mathbb R}$ \Comment{Pr(compete)}
\Require $\sigma \in [0.0, 1.0] \subset {\mathbb R}$ \Comment{Pr(reproduce)}
\Require $N_s \geq 3 \in {\cal Z}$ \Comment{Number of species}
\Require $S_{\text max} \geq 1 \in {\mathbb Z}$ \Comment{Max.\ \#MCS}
\Require $E_{\text max} \geq 1 \in {\mathbb Z}$ \Comment{Max.\ \#elementary-steps}
\Require ${\cal N} \in \{ \text{`vonNeumann', `Moore'} \} $ \Comment{Nbrhood spec}
\Require ${\cal B} \in \{\text{`periodic'}, \text{`noflux'}\} $ \Comment{Boundary condition}
\Ensure $N_s = 2j + 1 ; j \in {\mathbb Z}$ \Comment{$N_s$ must be odd}
\State $N \gets L^2$ \Comment{Total number of cells in lattice}
\Ensure $M \leq \frac{1}{2N}$ \Comment{S.t. $\epsilon \in [0.0,1.0] \in {\mathbb R}$ }
\State $\epsilon \gets 2MN$ \Comment{Pr(move)}
\State$x \gets 0$
\While{$x < L$}\Comment{Populate lattice with species}
	\State$y \gets 0$
	\While{$y<L$}	
		\State $l(x,y) \gets {\cal U}\{0, \ldots, N_s-1\}$
		\State$y\gets y +1  $
	\EndWhile
	\State$x\gets x +1 $
\EndWhile
\State $s \gets 0$ \Comment{$s$ is current MCS}
\While{$s < S_{\text max}$}\Comment{Outer MCS loop}
\State $e \gets 0$\Comment{$e$ is current elementary step}
\While{$e < E_{\text max}$}
	\Comment{Core inner ES loop}
	\State$\vec{p}_i \gets ({\cal U}\{0,\ldots, L-1\} ,  {\cal U}\{0,\ldots, L-1\} )$	
	\State$\vec{p}_n \gets {\text{\sc RndNeighbor}}(l, \vec{p}_i, {\cal N}, {\cal B}) $
	\State$l \gets$ {\sc ElStep}$(l, \vec{p}_i, \vec{p}_n, \mu, \sigma, \epsilon)$
\State $e \gets e+1$
\EndWhile 
\State $s \gets s+1$
\EndWhile
\end{algorithmic}
\end{algorithm}

In each individual ESCG experiment, the density $\rho_i(t)$ of each species $S_i$ (i.e., what proportion of the lattice cells are occupied by agents of species type $S_i$) was recorded after each MCS: illustrative time-series of the $\rho_i$ values from several experiments are shown below. The variation in densities at any one time is also relevant to the discussion later in this paper: for this the mean density at time $t$ is calculated as $\hat{\rho}(t)=\frac{1}{N_S}\sum^{N_S}_1 \rho_i(t)$ and then the variation in density $\rho_v(t)$ was calculated as the standard deviation around the mean:
$\rho_v(t)=(\frac{1}{N_S}{\sum^{N_S}_1 (\rho_i(t)-\hat{\rho}(t))^2})^{0.5}$.

\begin{figure}[h]
\begin{center}
\includegraphics[trim=0cm 0cm 0cm 0cm, clip=true, scale=0.55]{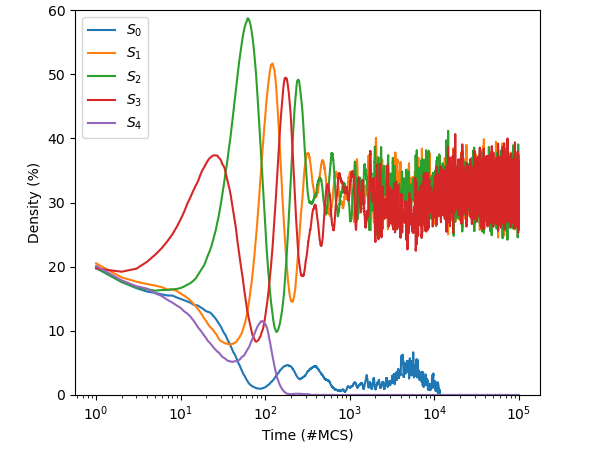}
\end{center}
\caption{
OES-based replication of Zhong et al.'s experiment for $L$$=$$200$, $\mu$$=$$\sigma$$=$$1.0$, $M$$=$$10^{-5}$, and with a single directed edge ablated from the dominance digraph (denoted by $N_a$$=$$1$). 
Plot shows time-series of the densities of five species labelled $S_0$ to $S_4$. In this experiment the ablated edge was 
$S_0$$\rightarrow$$S_3$. 
The end-state of the system after 100kMCS is three-species coexistence, denoted by $n_s(100\text{k})$$=$$3$. 
The horizontal axis is time, measured in MCS; and the vertical axis is density, expressed as a percentage: if at time $t$ the total headcount of species $S_i$ is $h_i(t)$ then $S_i$'s density 
is $\rho_i(t)$$=$$h_i(t) / L^2$. 
Here, peak density is $\rho_2(t)$$\approx$$59$ at $t$$\approx$$50$MCS.
}
\label{fig:Zhong_replicate_3species}
\end{figure}

\begin{figure}[h]
\begin{center}
\includegraphics[trim=0cm 0cm 0cm 0cm, clip=true, scale=0.55]{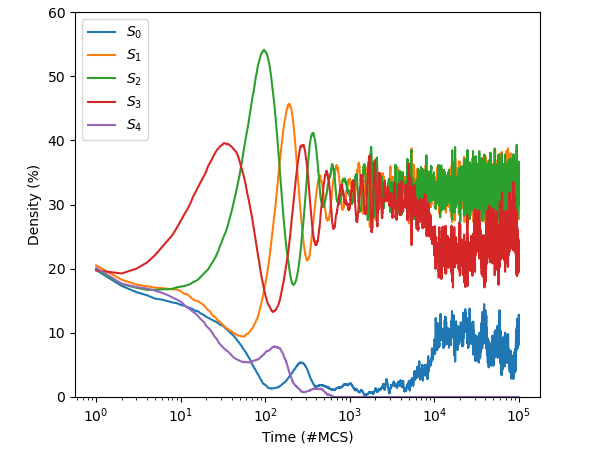}
\end{center}
\caption{OES-based replication of Zhong et al.'s experiment for $L$$=$$200$, $\mu$$=$$\sigma$$=$$1.0$, $M$$=$$10^{-7}$, and with the single directed edge $S_0$$\rightarrow$$S_3$ ablated from the dominance digraph (i.e., $N_a$$=$$1$). In this experiment the end-state of the  system after 100kMCS is four-species coexistence, denoted by $n_s(100\text{k})$$=$$4$. 
Format as for Figure~\ref{fig:Zhong_replicate_3species}. 
Peak density is $\rho_1(t)$$\approx$$55$ at $t$$\approx$$100$MCS.
}
\label{fig:Zhong_replicate_4species}
\end{figure}

\begin{figure}[t]
\begin{center}
\includegraphics[trim=0cm 0cm 0cm 0cm, clip=true, scale=0.35]{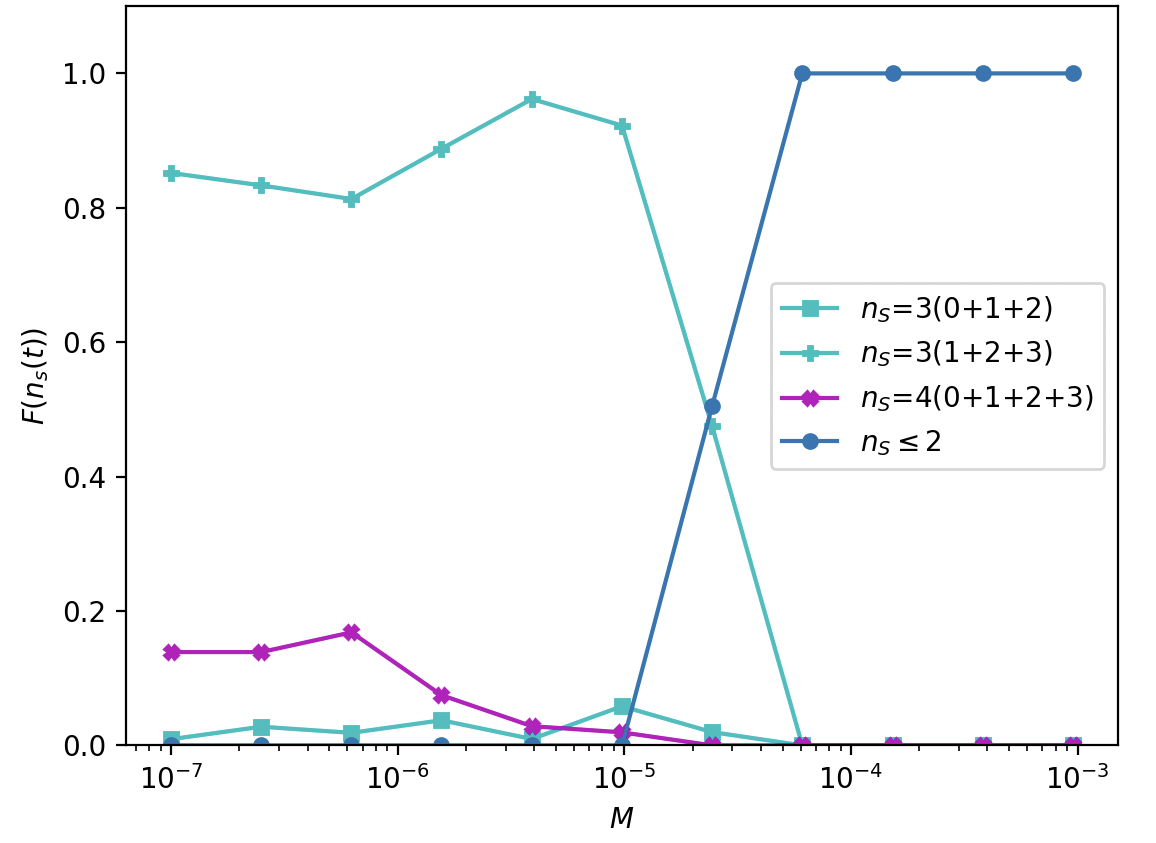}
\end{center}
\caption{
Occurrence frequency of possible ``asymptotic states'' species counts after 100000MCS (i.e., $n_s(100\text{k})$) in OES-based replication of Zhong et al.'s $N_a$$=$$1$ experiments, where species $S_0$ no longer dominates species $S_3$, with $L$$=$$200$, $\mu$$=$$\sigma$$=$$1.0$, $M$$\in$$[10^{-7}, 10^{-3}]$. 
Horizontal axis is mobility $M$; vertical axis is frequency of occurrence, denoted $F(n_s(t))$. 
For ease of comparison to Zhong et al.'s Figure~3, this plot follows Zhong et al.'s convention of aggregating the results for $n_s$$=$$1$ and $n_s$$=$$2$, but treating the two $n_s$$=$$3$ outcomes and the one  $n_s$$=$$4$ outcome individually. 
The legend shows, for each individual $F(n_s)$ outcome, the value of $n_s(t)$ followed in parentheses by the species-numbers of those species that survived to time $t$. 
}
\label{fig:Zhong_replicate_FvM_A}
\end{figure}

Results from individual simulations with my implementation\footnote{The Python code I wrote for the experiments in this paper is freely available on GitHub, released under the MIT Open-Source Licence. See: {\tt https://github.com/davecliff/ESCG\_Python}.} of OES (Algorithm~\ref{alg:OES}) being called from the ESCG (Algorithm~\ref{alg:Game}) and with ablated dominance digraphs, are in very good agreement with the results of \cite{zhong_etal_2022_ablatedRPSLS}, as is demonstrated by Figures~\ref{fig:Zhong_replicate_3species} 
to~\ref{fig:Zhong_replicate_FvM_A}.

Figure~\ref{fig:Zhong_replicate_3species} shows time-series results from a single run of the simulation where one directed edge is ablated from the dominance network (which I denote by $N_a$$=$$1$, for number of ablations), and the system's end state after 100kMCS is three-species coexistence (which I denote by $n_s(100\text{k})$$=$$3$, for number of species at time $t$$=$$100$kMCS).

Figure~\ref{fig:Zhong_replicate_4species} shows another run with  $N_a$$=$$1$, but in this case the end state is $n_s(100\text{k})$$=$$4$.  
These two figures are directly comparable to Zhong et al.'s Figures~2a and 2c, which are qualitatively indistinguishable from my results presented here. Zhong et al.\ use the phrase ``asymptotic state'' to refer to the end-state of the ESCG after 100kMCS, but, as I argue later in this paper, that is potentially misleading because the OES RPSLS system can exhibit transients much longer than 100kMCS, and may take ten times as long to settle on an asymptote. 

Similarly, results aggregated over multiple IID simulations with my implementation of OES being called from the ESCG are also in very good agreement with those published by \cite{zhong_etal_2022_ablatedRPSLS}. For example, Figure~\ref{fig:Zhong_replicate_FvM_A} shows the frequency of occurrence of possible ``asymptotic states'' in my OES-based ESCG when a single edge is ablated from the dominance network, such that species $S_0$ does not dominate species $S_3$: this is directly comparable to Zhong et al.'s Figure~3.
For ease of comparison, my Figure~\ref{fig:Zhong_replicate_FvM_A} follows Zhong et al.'s Figure~3 in grouping all $n_s(100\text{k})$$\leq$$2$ results together, but treating different $n_s(100\text{k})$$=$$3$ outcomes separately: Zhong et al.\ show results from ESCG simulations where the Rock-Scissors link is ablated (i.e., $N_a$$=$$1$), with separate traces for the two $n_s(100\text{k})$$=$$3$ outcomes Rock-Lizard-Spock and Scissors-Lizard-Spock, and another separate trace for the $n_s(100\text{k})$$=$$4$ outcome Rock-Scissors-Lizard-Spock). 
The results in Figure~\ref{fig:Zhong_replicate_FvM_A} are qualitatively indistinguishable from those in Figure~3 of \cite{zhong_etal_2022_ablatedRPSLS}.

However, the results shown in my Figure~\ref{fig:Zhong_replicate_FvM_A} and Zhong et al.'s Figure~3 only tell half the story. In the unablated dominance network, $S_0$ dominates both $S_1$ and $S_3$: my Figure~\ref{fig:Zhong_replicate_FvM_A} and Zhong et al.'s Figure~3 show what happens when the $S_0$$\rightarrow$$S_3$ dominance edge is ablated, but what happens when instead the   $S_0$$\rightarrow$$S_1$ edge is ablated? 
This $N_a$$=$$1$ case is not discussed anywhere in \cite{zhong_etal_2022_ablatedRPSLS}, and they offer no guidance on why they did not explore this case. 

\begin{figure}[h]
\begin{center}
\includegraphics[trim=0cm 0cm 0cm 0cm, clip=true, scale=0.55]{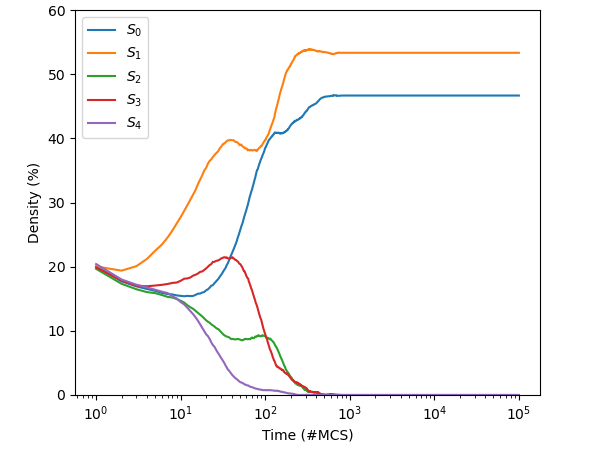}
\end{center}
\caption{OES-based exploration of  Zhong et al.'s experiment for $L$$=$$200$, $\mu$$=$$\sigma$$=$$1.0$, $M$$=$$10^{-7}$, and with the single directed edge ablated from the dominance digraph ($N_a$$=$$1$) being $S_0$$\rightarrow$$S_1$. In this run of the simulation, $n_s(100\text{k})$$=$$2$. 
Format as for Figure~\ref{fig:Zhong_replicate_3species}. 
}
\label{fig:Zhong_replicate_2_species}
\end{figure}

Figure~\ref{fig:Zhong_replicate_2_species} shows results from a single simulation when the $N_a$$=$$1$ ablation is of $S_0$$\rightarrow$$S_1$, and the outcome is $n_s(100\text{k})$$=$$2$; for this ablation,  $n_s(100\text{k})$$=$$2$ is the outcome for the entire range of $M$ values sampled. Presumably Zhong et al.\ would also have seen this result if they had thought to run this complementary set of simulations. As can be seen, this case does {\em not} show a sudden collapse in the occurrence frequency of $F(n_s(100\text{k}))$$=$$3$ as $M$ is increased past some threshold value: instead the outcome is constant. Zhong et al.'s observation of the sudden steep drop in $n_s$=$3$ outcomes does hold true when the ablated edge is $S_0$$\rightarrow$$S_3$, but it does not when the ablation is $S_0$$\rightarrow$$S_1$.

\begin{figure}
\begin{center}
\includegraphics[trim=0cm 0cm 0cm 0cm, clip=true, scale=0.33]{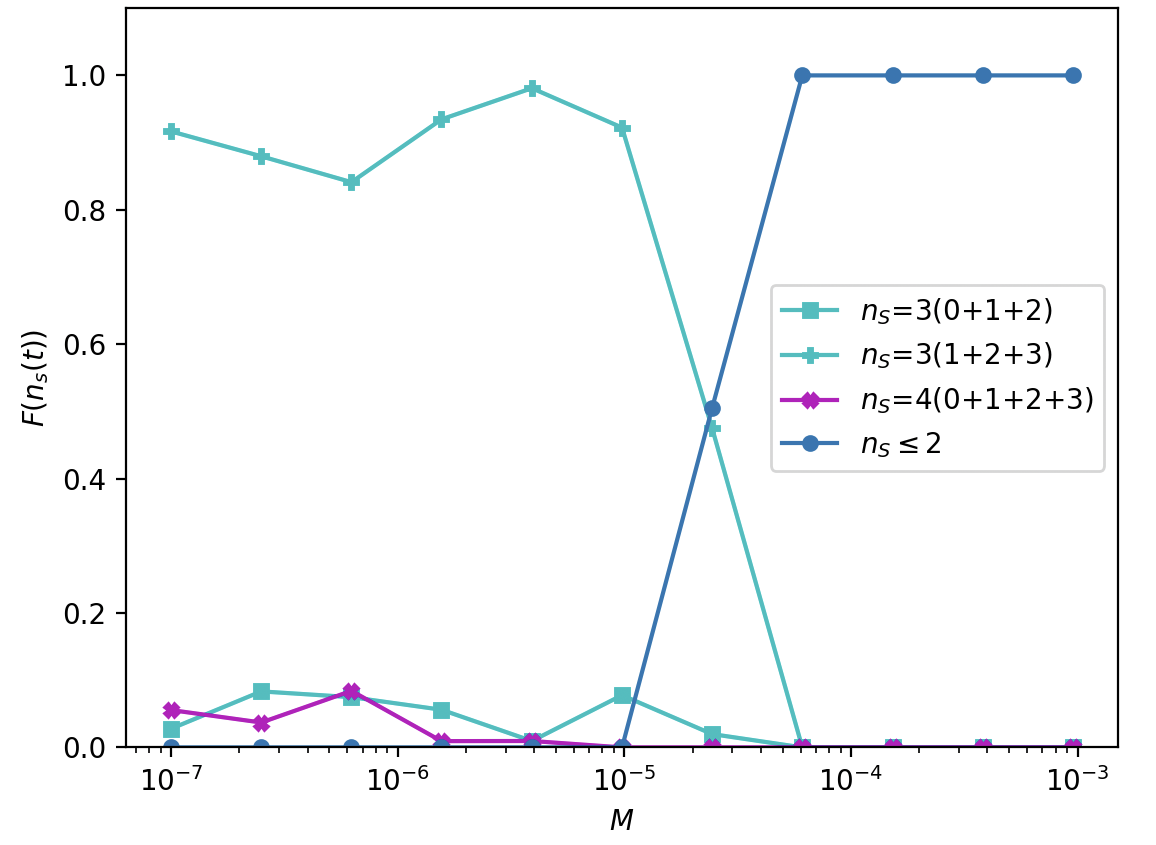}
\end{center}
\caption{
Occurrence frequency $F(n_s(200k))$ after 200kMCS, twice as long as was shown in Figure~\ref{fig:Zhong_replicate_FvM_A}, for
OES-based replication of Zhong et al.'s $N_a$$=$$1$ experiments, with $L$$=$$200$, $\mu$$=$$\sigma$$=$$1.0$, $M$$\in$$[10^{-7}, 10^{-3}]$, 
in which the single ablation deletes $S_0$'s domination of $S_3$. 
Format as for Figure~\ref{fig:Zhong_replicate_FvM_A}.
}
\label{fig:Zhong_replicate_FvM_B}
\end{figure}

Finally, bringing this replication and exploration of  \cite{zhong_etal_2022_ablatedRPSLS} to a close, Figure~\ref{fig:Zhong_replicate_FvM_B} shows the occurrence frequencies after 200kMCS, i.e.\ twice as long as Zhong et al.\ ran their experiments for, and twice as long as the results shown in Figure~\ref{fig:Zhong_replicate_FvM_A}. Comparing  Figure~\ref{fig:Zhong_replicate_FvM_B} to Figure~\ref{fig:Zhong_replicate_FvM_A}, it is clear that the frequency of $n_s(t)$$=$$4$ outcomes drops significantly when given twice as long to run (and consequently there is an increase in the frequency of  $n_s(t)$$=$$3$ outcomes with species $S_0, S_1$, and $S_2$ remaining), and hence that Zhong et al.'s repeated reference to their $F(n_s(100\text{k}))$ results as the ``asymptotic state'' is potentially misleading, because the system clearly has not come close to an asymptote at $t$$=$$100$kMCS: presumably, if run for even longer (e.g.\ out to $t$=$10^6$MCS), the frequency of $n_s(t)$$=$$4$ outcomes would fall to zero (or very near to zero). 

\subsection{Critique of the Original Elementary Step (OES) }

Note that the OES only makes one call to the simulation platform's pseudo-random number generator (PRNG): the assignment of a uniform random real value in $[0.0,1.0]$ to $a$ at line 4, and the value of $a$ is then used to select which one of the three possible actions are executed. However, once the algorithm is expressed in pseudocode, it is easier to recognise there are several scenarios where the call to OES achieves absolutely nothing, and so the call to the PRNG is wasted computational effort: a random number is generated, but then the specific contents of $c_i$ and $c_n$ can mean that no changes are made to the lattice. In the parlance of low-level assembly-language and machine-code programming, an instruction that achieves absolutely nothing, other than wasting the processor-time it takes to execute, is often known as a {\em no-op} (from ``no operation'', because sometimes it is necessary to require a central processor unit to waste one or more clock cycles without doing anything else, to ensure timing and synchronization constraints are met). The situations in which a call to OES is effectively a no-op  are as follows:
\begin{itemize}
\item If $c_i$$=$$\emptyset$ and $c_n$$=$$\emptyset$, the call to {\sc ElStep} will be a no-op.
\item If $c_i$$=$$\emptyset$ or $c_n$$=$$\emptyset$, the call to {\sc Compete} will be a no-op. 
\item If $c_i$$\neq$$\emptyset$ and $c_n$$\neq$$\emptyset$, the call to {\sc Reproduce} will be a no-op.
\end{itemize} 
Empirical testing with my implementation of Zhong et al.'s algorithm revealed that, depending on circumstances, as many as 30\% of calls to {\sc ElStep} are no-ops. So potentially of the 160 trillion calls to ES in generating results for Zhong et al.'s paper, roughly 48 trillion calls to the PRNG were made within {\sc ElStep} but had no effect because they were under no-op conditions. If each call to the PRNG takes one microsecond\footnote{On a 2020-vintage Apple Mac Mini with M1 silicon, using Python's {\tt timeit} on {\tt random.random()} indicates that a single call to the {\tt random} PRNG algorithm takes roughly $10^{-6}$ seconds.} then the time spent in generating random numbers for no-op conditions is 4.8~million seconds, which is 1333 hours, or 55 days of real wall-time computation, all wasted.  
 
\begin{figure}
\begin{center}
\includegraphics[trim=0cm 0cm 0cm 0cm, clip=true, scale=0.55]{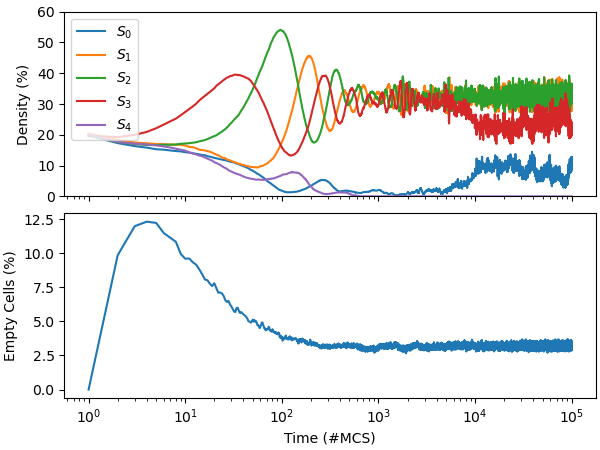}
\end{center}
\caption{Percentage of empty cells in the lattice over the course of the $L$$=$$200$ OES experiment shown in Figure~\ref{fig:Zhong_replicate_3species}. Peak empty-cell count occurs at $t$$\approx$$5$MCS and is roughly 12\% of the total lattice area, and the count of empty cells does not reach a stable dynamic equilibrium until $t$$\approx$$100$MCS.
}
\label{fig:n_empty}
\end{figure}

Furthermore, because of the (unexplained, unjustified) commitment to only ever executing one of the three possible actions per ES, calls to {\sc Compete} can leave one of the two cells empty, but subsequent calls to {\sc Reproduce} have no record of any recently-created empty cells that could be bred into. In cases where (as in the ESCG of Algorithm~\ref{alg:Game}) the initialisation of the lattice fills every cell with an individual, initially any call to OES that attempts a {\sc Reproduce} will fail, because there are no empty cells. Over time the number of empty cells will increase as ever more individuals are killed in calls to {\sc Compete}, and as empty cells become more prevalent so calls to {\sc Reproduce} are ever more likely to succeed, with each successful reproduction reducing the count of empty cells by one. Eventually the system will settle to a stable dynamic equilibrium, with many cells being emptied by competition and at the same time many cells re-filled by reproduction, with the two forces balancing each other out. To illustrate this, Figure~\ref{fig:n_empty} shows the count of empty cells over the duration of the experiment whose population dynamics were shown in Figure~\ref{fig:Zhong_replicate_3species}: in this experiment, the eventual steady-state dynamic equilibrium is for roughly 3\% of the cells to be empty at any one time, but in other circumstances the empty-cell count can hold at $>$$10\%$ for extended periods.  
Although the model having empty cells
might seem to increase 
biological plausibility, my view is that these ESCG models are so minimal, so abstract, that spending 3\% or more of the available space modelling nothing just wastes valuable resources, because more empty cells in the grid just means more calls to {\sc ElStep} and {\sc Compete} that will be no-ops: the empty cells simply do not add anything of value to the model's dynamics.

These considerations led me to rewrite the OES, and my Revised Elementary Step (RES) algorithm is introduced in the next section. 

\subsection{Revised Elementary Step (RES)}
\label{sec:RES}

Algorithm~\ref{alg:RES} shows the RES in its entirety. All of the revisions have been introduced to address the points of criticism raised in the previous section. The RES version of {\sc ElStep} commences with a check to see if both cells are empty: if they are, it immediately returns with no changes made to the lattice, but also without any invocation of the PRNG.  It then checks to see if both cells contain an individual: if they do, it invokes the PRNG at line 9 and if the random value is less than $\mu$ it calls {\sc Compete} at line 10, updating the lattice accordingly. This means that by the time execution passes to line 14 we know that we either have two individuals because either they didn't compete or they did but the outcome was a draw, or instead one of the two cells is empty either because it was passed into {\sc ElStep} as an empty cell, or it was emptied by the immediately preceding call to {\sc Compete} at line~10; there is no way we can get to line~14 and both cells be empty (because we would have caught that at line~4) so if either cell is empty that means we can invoke the PRNG at line~15 and if the resulting random value is less than $\sigma$ we call {\sc Reproduce}, thereby filling the empty cell. Finally, regardless of the contents of either of the two cells, we make a final call to the PRNG at line~20 and if that random value is less than $\epsilon$ we invoke {\sc Move} to swap the contents of the two cells.

\begin{algorithm}[t]
\caption{Revised Elementary Step (RES)}
\label{alg:RES}
\begin{algorithmic}[1]
\Require $\mu \in [0.0, 1.0] \subset {\mathbb R}$ \Comment{Pr(compete)}
\Require $\sigma \in [0.0, 1.0] \subset {\mathbb R}$ \Comment{Pr(reproduce)}
\Require $\epsilon \in [0.0, 1.0] \subset {\mathbb R}$ \Comment{Pr(move)}
\Require $l$ \Comment{Current state of lattice $l$}
\Require $\vec{p}_i$ \Comment{Lattice coords of cell $c_i$}
\Require $\vec{p}_n$ \Comment{Lattice coords of cell $c_n$}
\Procedure{ElStep}{$l, vec{p}_i, vec{p}_n, \mu, \sigma, \epsilon$} 

	\State
	\State \Comment{if both cells empty, do nothing}
	\If{$ l(\vec{p}_i) = \emptyset$ \textbf{and}  $l(\vec{p}_n) = \emptyset$}
		\State \textbf{return} $l$
	\EndIf
	\State \Comment{if both cells occupied, can compete}
	\If{$ l(\vec{p}_i) \neq \emptyset$ \textbf{and}  $l(\vec{p}_n) \neq \emptyset$}
		\If {${\cal U}\left[0.0,1.0\right] \leq \mu$}
    			\State $l  \gets ${\sc Compete}$(l, \vec{p}_i, \vec{p}_n)$
		\EndIf	
	\EndIf
	\State \Comment{if either cell empty, can reproduce}
	\If{$ l(\vec{p}_i) = \emptyset$ \textbf{or}  $l(\vec{p}_n) = \emptyset$}
		\If{${\cal U}\left[0.0, 1.0\right] \leq \sigma$}
			\State $l \gets ${\sc Reproduce}$(l, \vec{p}_i, \vec{p}_n)$
		\EndIf
	\EndIf
	\State \Comment{finally move}
	\If {${\cal U}\left[0.0, 1.0\right] \leq \epsilon$}
    		\State $l \gets ${\sc Move}$(l, \vec{p}_i, \vec{p}_n)$
	\EndIf    
	
	\State \textbf{return} {$l$}
\EndProcedure
\end{algorithmic}
\end{algorithm}

Note that, given the way RES is structured, in the limit-case of $\sigma$$=$$1.0$ (as used by \cite{zhong_etal_2022_ablatedRPSLS}, and many other authors), the number of empty cells will be forever zero: every time an empty cell is created in the call to {\sc compete} at line 10, that empty cell is then immediately filled by the call to {\sc Reproduce} at line 16: this entirely eliminates the space inefficiency in OES.  Also, in RES, the probabilities $\mu, \sigma$, and $\epsilon$ really are treated as literal probabilities: for example, $\mu$$=$$0.5$ really does mean a 50\% chance of competing on each call to the RES version of {\sc ElStep}, in contrast to the ``normalizable'' interpretation of the values of these parameters in OES where, for instance $\mu$$=$$0.5$ might actually mean a 100\% chance of competing (i.e., if $\epsilon$$=$$\sigma$$=$$0.0$) or a 20\% chance of competing (i.e., if  $\epsilon$$=$$\sigma$$=$$1.0$) or any other probability between 0.2 and 1.0 depending on the values assigned to $\epsilon$ and  $\sigma$ .

Note also that the upper-bound on the number of cell-to-cell (inter-)actions  per call to {\sc ElStep} is three in RES, but only one in OES. In those terms then, when using RES, we can expect the evolutionary dynamics to unfold at up to three times the pace of otherwise comparable OES experiments. This means the MCS-counts on the sequences of events will be compressed in RES experiments, relative to the same counts in OES experiments. Furthermore, when we factor in the potential for say 30\% no-op rate in OES, the peak rate of three actual (not no-op) actions per ES in RES vs.\ an average of roughly 0.7 in OES means the OES:RES speedup coefficient on events per MCS could in principle be as high as ($3/0.7$$\approx$)4.3.  
 
However, anyone with a keen eye for performance optimization will recognise in Algorithm~\ref{alg:RES} that the number of invocations  of the PRNG in any one call to the RES version of {\sc ElStep} can vary between zero and three, versus a  constant one PRNG call per OES {\sc ElStep}. Given that invocation of the PRNG is by far the most computationally intense operation within the entire ES, it is not impossible that actually, despite RES introducing checks to eliminate the no-op cases, the additional calls to the PRNG made by the RES {\sc ElStep} will cause the actual RES overall experiment execution wall-clock times to be slower than those for OES. Whether this happens in practice or not is explored next.

\section{Results}
\label{sec:results} 

\subsection{Results from Ablated Dominance Network, $N_a$$=$$1$}
\label{sec:resultsablated}

\begin{figure}[t]
\begin{center}
\includegraphics[trim=0cm 0cm 0cm 0cm, clip=true, scale=0.55]{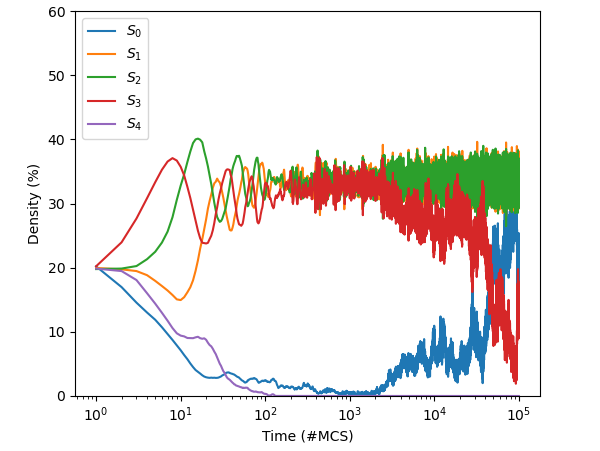}
\end{center}
\caption{RES-based replication of Zhong et al.'s experiment for $L$$=$$200$, $\mu$$=$$\sigma$$=$$1.0$, $M$$=$$10^{-7}$, with the  single directed edge $S_0$$\rightarrow$$S_3$ ablated from the dominance digraph (i.e., $N_a$$=$$1$). In this simulation run the end-state is $n_s(100\text{k})$$=$$4$, i.e.\ four-species coexistence. Format as for Figure~\ref{fig:Zhong_replicate_4species}. 
Here, peak density in the RES system is $\rho_2(11)$$\approx$$40$, roughly 30\% less than the comparable peak in the OES system shown in Figure~\ref{fig:Zhong_replicate_4species}. 
}
\label{fig:Zhong_replicate_4species_RES}
\end{figure}

\begin{figure} [h]
\begin{center}
\includegraphics[trim=0cm 0cm 0cm 0cm, clip=true, scale=0.55]{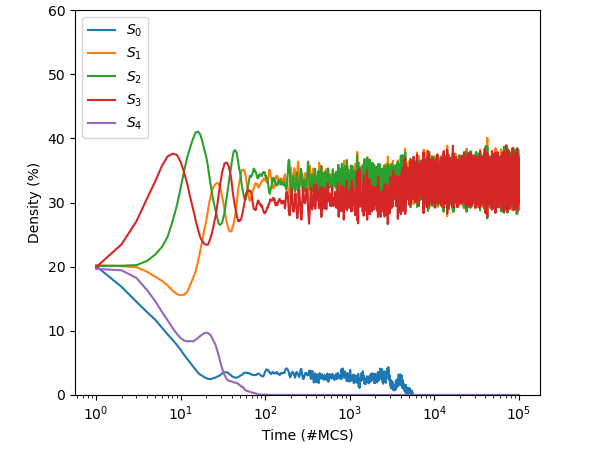}
\end{center}
\caption{RES-based replication of Zhong et al.'s experiment for $L$$=$$200$, $\mu$$=$$\sigma$$=$$1.0$, $M$$=$$10^{-7}$, with the  single directed edge $S_0$$\rightarrow$$S_3$ ablated from the dominance digraph (i.e., $N_a$$=$$1$). In this simulation run the end-state is $n_s(100\text{k})$$=$$3$, i.e.\ three-species coexistence. Format as for Figure~\ref{fig:Zhong_replicate_4species}. 
Here again, peak density in the RES system is $\rho_2(11)$$\approx$$40$, roughly 25\% less than the comparable peak in the OES system shown in Figure~\ref{fig:Zhong_replicate_3species}. 
}
\label{fig:Zhong_replicate_3species_RES}
\end{figure}

\begin{figure} [h]
\begin{center}
\includegraphics[trim=0cm 0cm 0cm 0cm, clip=true, scale=0.55]{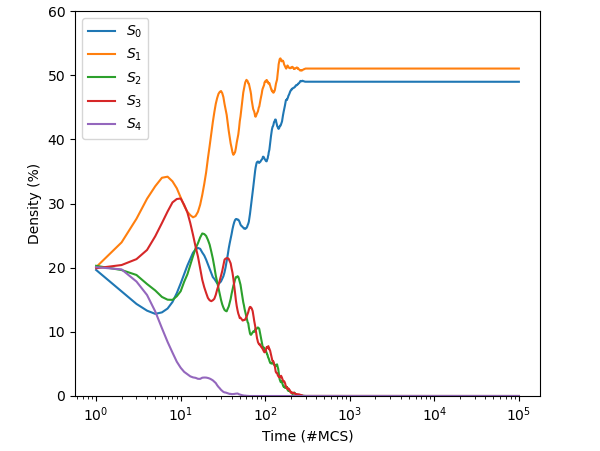}
\end{center}
\caption{RES-based extension of Zhong et al.'s experiment for $L$$=$$200$, $\mu$$=$$\sigma$$=$$1.0$, $M$$=$$10^{-7}$, here with the  single directed edge $S_0$$\rightarrow$$S_1$ ablated from the dominance digraph (i.e., $N_a$$=$$1$), an ablation not explored by Zhong et al.\ In this simulation run the end-state is $n_s(100\text{k})$$=$$2$, i.e.\ two-species coexistence. Format as for Figure~\ref{fig:Zhong_replicate_4species}. 
}
\label{fig:Zhong_extend_3species_RES}
\end{figure}

Figures~\ref{fig:Zhong_replicate_4species_RES}  to~\ref{fig:Zhong_extend_3species_RES} show results from $L$$=$$200$ $N_a$$=$$1$ ESCG experiments using RES, for comparison with the example OES-based results shown in Figures~\ref{fig:Zhong_replicate_4species} and~\ref{fig:Zhong_replicate_3species}, respectively. As can be seen, the results from these individual RES-based experiments are qualitatively very similar to the OES-based results. As predicted, the population dynamics with RES move faster than with OES: note, for instance, that the initial peak density (the highest point on the green line) in the OES experiments occurs around $t$$\approx$$45$MCS, whereas in the RES results it occurs around $t$$=$$7$MCS. 

Comparison of the wall-clock runtimes of otherwise identical OES and RES experiments revealed that RES does offer a saving in absolute runtimes: typically a RES experiment will run 18\% faster than the equivalent OES explanation. (Using an implementation of the RPSLS ESCG written in the C programming language, running on an Apple MacBook Pro M2, the average runtime for an OES simulation with $L$$=$$200$ over 200kMCS was 10:48, and the average runtime for the same simulation using RES was 8:51, a difference of 117s). This 18\% speedup is a lot less than the $4.3$$\times$ speedup that was speculated upon above: evidently, much of the time-saving resulting from the elimination of no-ops in OES is lost because of RES's more frequent use of the PRNG. However, as we shall see in Section~\ref{sec:unablated}, RES offers more evolutionarily stable dynamics, which allows for considerably smaller values of $L$ to be used in RES experiments, in comparison to OES.  

\begin{figure}[t]
\begin{center}
\includegraphics[trim=0cm 0cm 0cm 0cm, clip=true, scale=0.35]{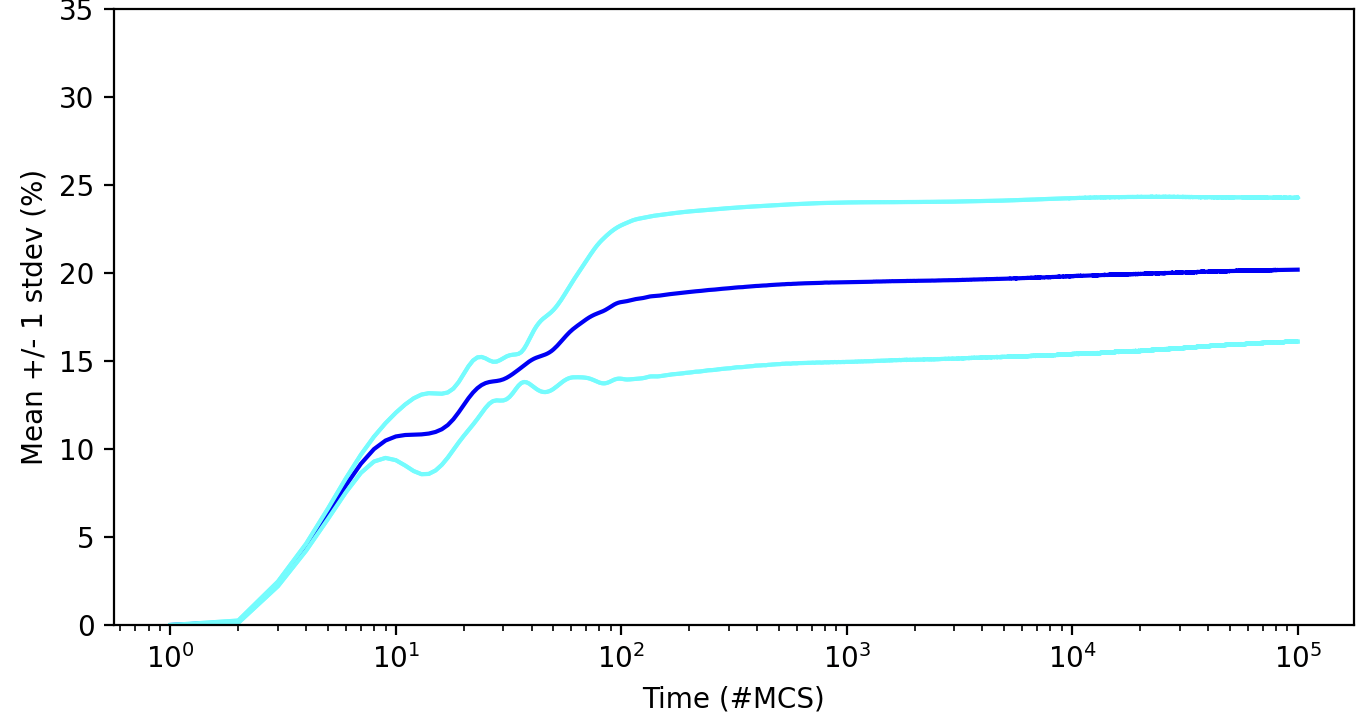}
\end{center}
\caption{Time series of mean (plus and minus one standard deviation) of variation in densities (denoted by $\rho_V(t)$)
across all five species in the 
  $N_a$$=$$1$ RES experiments with  $L$$=$$200$, $\mu$$=$$\sigma$$=$$1.0$, and $M$$\in$$[10^{-8},10^{-3}]$ . Horizontal axis is time, measured in MCS;  
vertical axis is $\rho_V(t)$, expressed as a percentage of $L^2$. At each value of $M$ sampled, 150 IID simulations were run.}
\label{fig:NS05_NA1_L200_sdevs}
\end{figure}

\begin{figure}
\begin{center}
\includegraphics[trim=0cm 0cm 0.0cm 0cm, clip=true, scale=0.35]{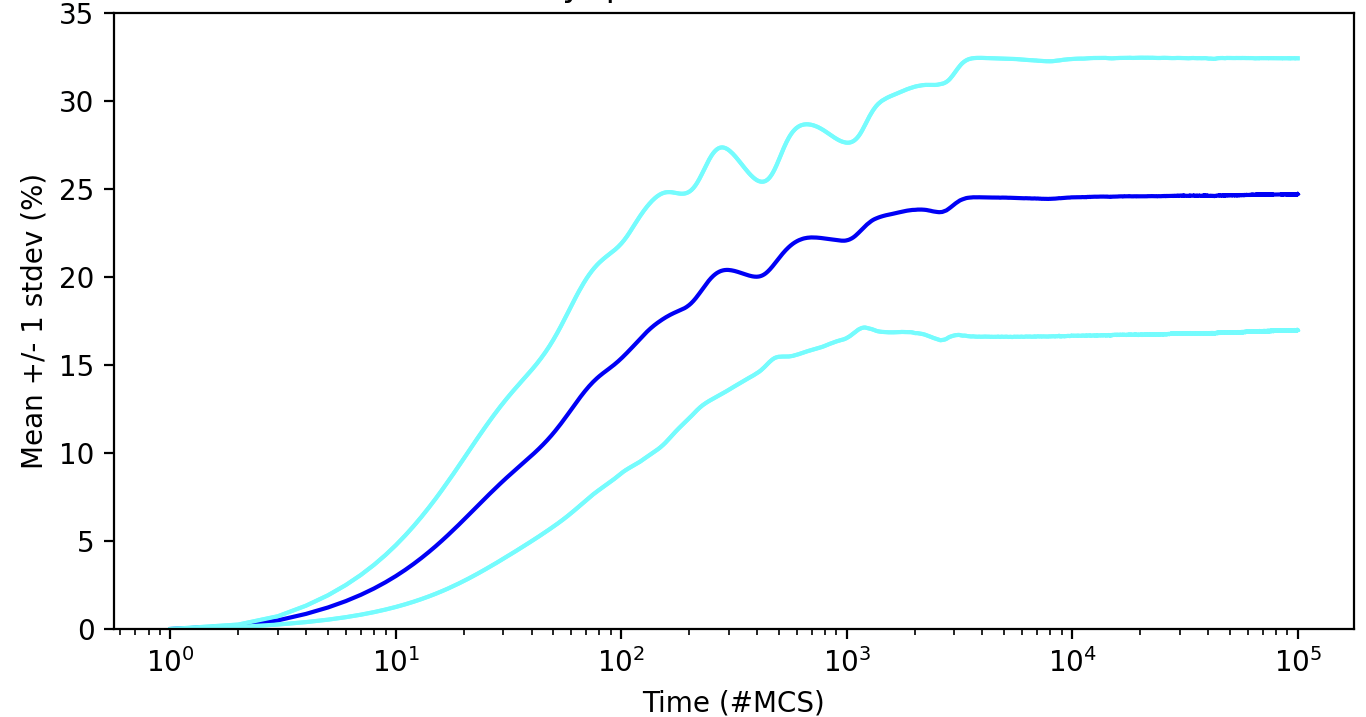}
\end{center}
\caption{Mean (plus and minus one standard deviation)  $\rho_V(t)$ from the five-species  $N_a$$=$$1$ OES experiments with  $L$$=$$200$, $\mu$$=$$\sigma$$=$$1.0$, and $M$$\in$$[10^{-8}, 10^{-3}]$.  Format as for Figure~\ref{fig:NS05_NA1_L200_sdevs}.
}
\label{fig:NS05_NA1_L200_Noop_sdevs}
\end{figure}

To summarise the system dynamics over a large number of IID experiments, Figure~\ref{fig:NS05_NA1_L200_sdevs} shows time series of the mean, plus and minus one standard deviation, of the variance in the population densities from 150 IID RES $N_a$$=$$1$ experiments over a range of mobility values $M$$\in$$[10^{-7}, 10^{-3}]$ and  
Figure~\ref{fig:NS05_NA1_L200_Noop_sdevs} shows a corresponding plot from 150 IID $N_a$$=$$1$ experiments with OES instead of RES. As can be seen, the species-density variance in the RES experiments settles to a steady state of $\approx$$20\%$$\pm$$5\%$ by $t$$\approx$$10^2$MCS, whereas the variance in the OES experiments doesn't settle until $t$$\approx$$10^4/2$MCS, and,  once it has settled, the OES steady-state variance is $\approx$$25\%$$\pm$$7\%$, considerably higher than the variance in the RES system. 

\subsection{Results from Unablated Dominance Networks}
\label{sec:unablated}

\cite{zhong_etal_2022_ablatedRPSLS} only reported results from RPSLS experiments with ablated dominance networks: their paper does not include any analysis or discussion of the dynamics of 2D ESCGs using the unablated RPSLP digraph. The results I present in this section provide a comparison between the population dynamics of OES and RES-based experiments in which the RPSLS dominance network is fully intact, i.e. where the number of ablations $N_a$$=$$0$.

\subsubsection{Revised Evolutionary Step (RES)}

\begin{figure}[t]
\begin{center}
\includegraphics[trim=0cm 0cm 0cm 0cm, clip=true, scale=0.35]{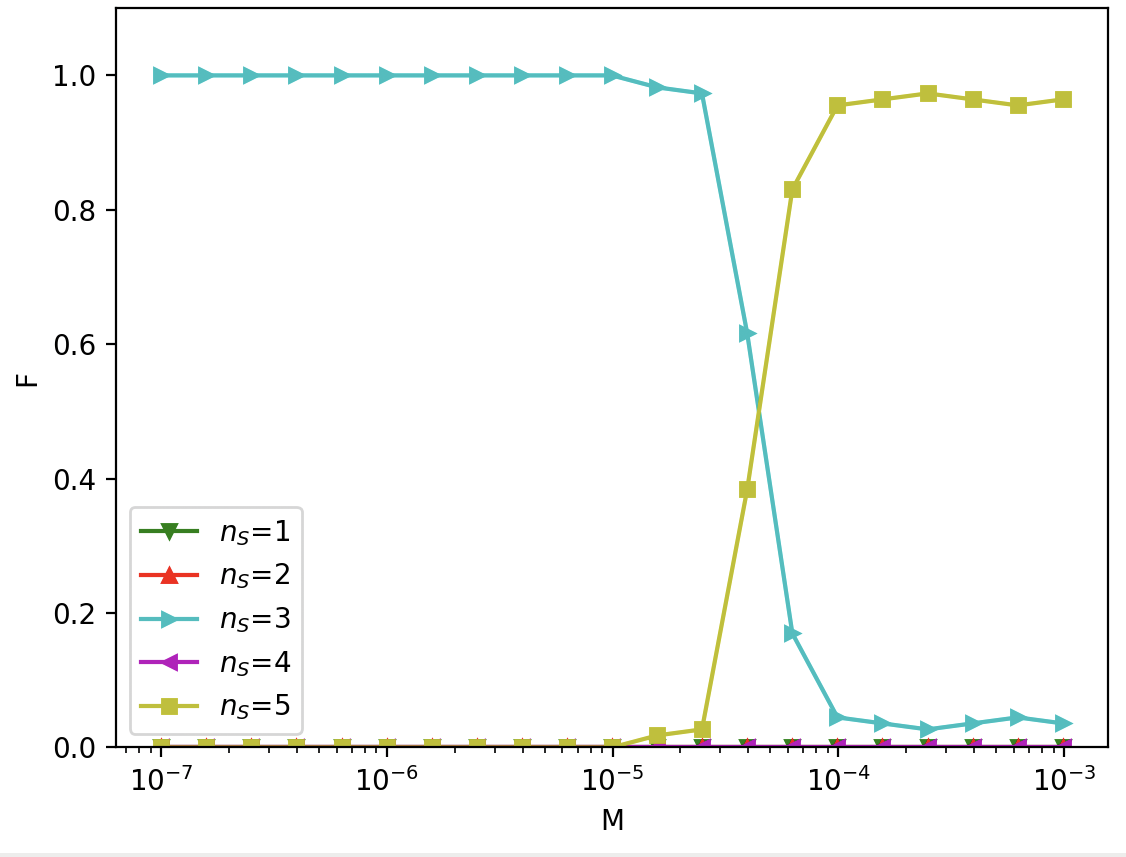}
\end{center}
\caption{Frequency of species-count outcome, denoted $F(n_s(t))$, from RES experiments with $N_a$$=$$0$ (no ablations) for $L$$=$$75$, $\mu$$=$$\sigma$$=$$1.0$, with 100 IID simulations at each value of $M$. Format as for Figure~\ref{fig:Zhong_replicate_FvM_A}. 
}
\label{fig:RES_NS05_NA0_L100_FvM}
\end{figure}

Figure~\ref{fig:RES_NS05_NA0_L100_FvM} shows $F(n_s(t))$ against $M$ at $t$$=$$200$kMCS  from RES experiments with $N_a$$=$$0$ (no ablations) for $L$$=$$75$, $\mu$$=$$\sigma$$=$$1.0$, with 100 IID simulations at each value of $M$. As can be seen,  for low values of $M$, the outcome is consistently $n_s(200k)$$=$$3$ (i.e., two of the five species have gone extinct) and this deserves some explanation because this is for the {\em unablated}\/ dominance network: if there are no ablations, given that the game is symmetric, {\em prima facie} one would reasonably expect the system dynamics to not produce any extinctions.

\begin{figure}[t]
\begin{center}
\includegraphics[trim=0cm 0cm 0cm 0cm, clip=true, scale=0.3]{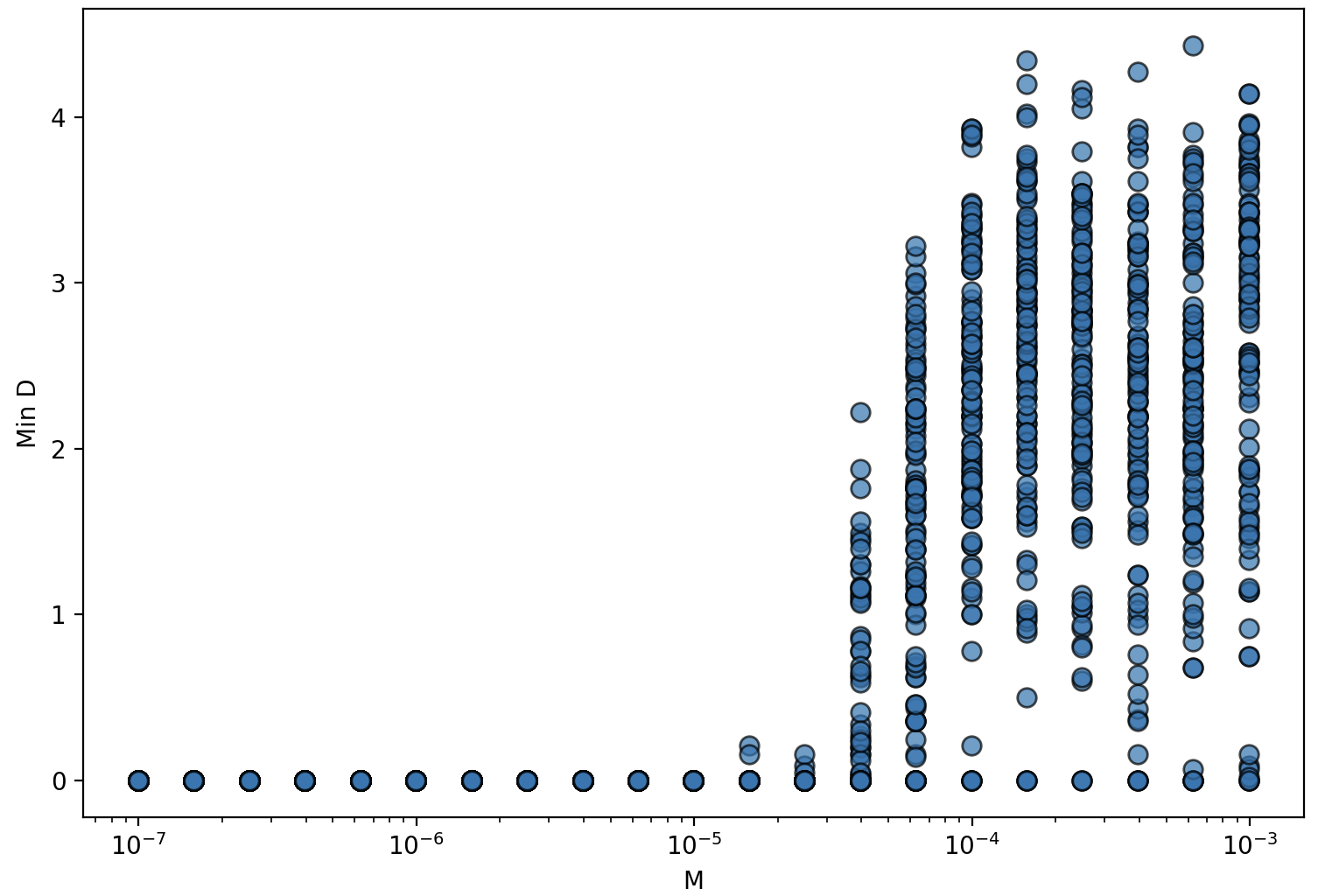}
\end{center}
\caption{Minimum densities after 200kMCS from RES experiments with $N_a$$=$$0$ (no ablations) for $L$$=$$75$, $\mu$$=$$\sigma$$=$$1.0$, with 100 IID simulations at each value of $M$$\in$$[ 10^{-7}, 10^{-3}]$.. Horizontal axis is $M$; vertical axis shows, for each of the 100 IID simulations at any one value of $M$, the minimum density (as a percentage) recorded for any species at any time in the whole simulation. 
The row of data-points at zero density reflect simulations where one or more species went extinct. Given that there are no exogenous causes of extinction in these simulations, and the dominance network is unablated, the cause of the extinctions is the choice of $L$$=$$75$, a lattice too small to contain the system's inherent fluctuations in species density, enabling ``underflow extinctions''; see text for further discussion.
}
\label{fig:RES_NS05_NA0_L100_minD}
\end{figure}

To better illustrate what is happening in this unablated ESCG, Figure~\ref{fig:RES_NS05_NA0_L100_minD} shows a scatter-plot of minimum densities in the 100 IID simulations illustrated in Figure~\ref{fig:RES_NS05_NA0_L100_FvM}: each data-point is the minimum density (as a percentage) recorded for any species in the ESCG at any time in the whole simulation, and at each value of $M$ there are 100 (often overlapping) data-points. The row of data-points at zero density reflect simulations where one or more species went extinct. Given that there are no exogenous causes of extinction in these simulations, and the dominance network is unablated, the  extinctions can only be due to the choice of $L$$=$$75$. That is, the lattice is simply too small to adequately contain the system's inherent fluctuations in species density. 

To explain this, imagine, for example, that there is an RPSLS-like ESCG in which we can analytically prove that the density of each species will fluctuate over time, and its minimum species density will never go below 0.1\%, for all time. If we use $L$$=$$200$, $N$$=$$L^{2}$$=$$40,000$, the minimum density in this hypothetical example would involve the headcount of a particular species falling to $40,000$$\times$$0.001$$=$$40$, before then rising again, and no extinctions occur. However, if we instead use $L$$=$$20$, then $N$$=$$L^{2}$$=$$400$, and the analysis indicates that the minimum species headcount will be 0.4, but the ESCG does not deal in fractional agents: the lattice is {\em discrete}, and a cell is either occupied by an agent or it is empty, and so the theoretical 0.4 headcount manifests itself in the ESCG simulation as a zero, an extinction -- and so there will be {\em endogenous} extinctions purely as a consequence of choosing too small a value for $L$. In computer science, if a nonzero floating-point number is too small to be representable in a particular data-type, and is instead represented as zero, that is referred to as {\em arithmetic underflow}, and I'll borrow that term here and refer to these endogenous extinctions caused by the lattice being too small as {\em underflow extinctions}.  

Note also that when an underflow extinction occurs and the headcount for some species $S_i$ goes to zero, that species is then extinct for the remainder of the simulation and is no longer an active participant in any RPSLS competitions, so the two other species that $S_i$ dominated now benefit from an absence of predation, while the two other species that dominated $S_i$ have now lost a source of prey -- the dominance digraph has lost four directed edges, i.e.\ the extinction of $S_i$ causes the system to go from $N_a$$=$$0$ to $N_a$$=$$4$. After the first such underflow extinction in a five-species RPSLS, the system might plausibly stabilise at four species, or the loss of the first species may trigger a domino-effect chain reaction of one or more further extinctions, resulting in the final asymptotic state species-count being fewer than four. 
Time series illustrating this domino effect, chains of extinctions caused by an initial underflow extinction, in OES $N_a$$=$$0$ ESCGs are shown in the appendices of the this paper.

As is shown in Figure~\ref{fig:RES_NS05_NA0_L150_minD}, the underflow extinction problem disappears once the lattice side-length $L$ is large enough to give sufficiently many cells that can accommodate the natural fluctuations in densities of the evolving populations, without the lattice discretization forcing extinctions, for whatever terminal MCS-count is being used to monitor the ``asymptotic state''. In technical terms, we need a value of $L$ sufficiently large that the system's {\em mean extinction time} (MET) is sufficiently long that extinctions are highly unlikely before the end of the simulation, such that the expected value of $n_s(t_{\text max})$ is essentially the same as $N_s$. For further discussion of MET in ESCGs, see \cite{viswanathan_etal_2024}

\begin{figure}[t]
\begin{center}
\includegraphics[trim=0cm 0cm 0cm 0cm, clip=true, scale=0.3]{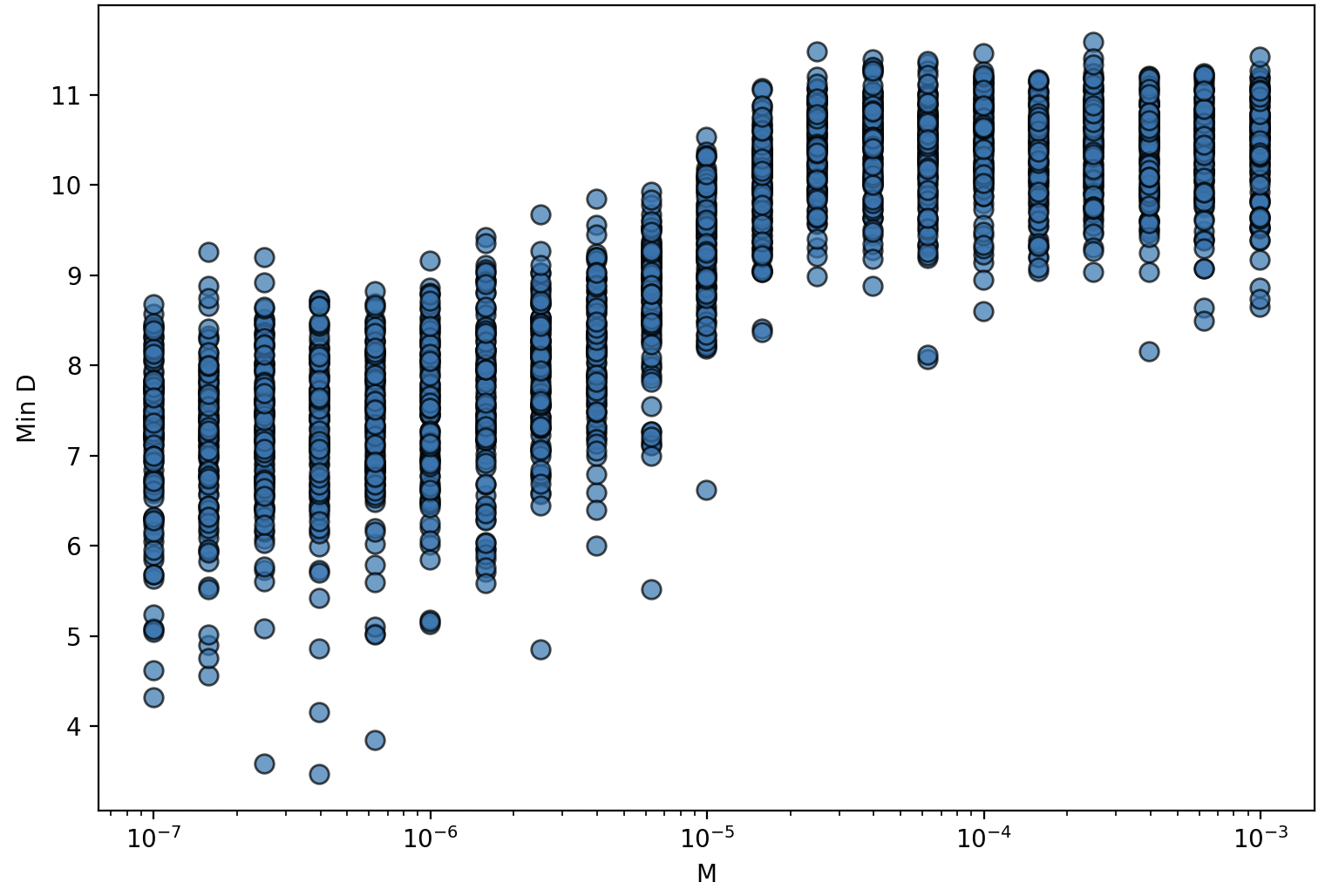}
\end{center}
\caption{Minimum density after 200kMCS from RES experiments with $N_a$$=$$0$ (no ablations) for $L$$=$$150$, $\mu$$=$$\sigma$$=$$1.0$, with 100 IID simulations at each value of $M$$\in$$[ 10^{-7}, 10^{-3}]$.. Format as for Figure~\ref{fig:RES_NS05_NA0_L100_minD}. With the increased lattice size ($N$ here is four times the $N$ of Figure~\ref{fig:RES_NS05_NA0_L100_minD}), the minimum density is always above 2\% and no underflow extinctions occur;  see text for further discussion.
}
\label{fig:RES_NS05_NA0_L150_minD}
\end{figure}

\begin{figure}
\begin{center}
\includegraphics[trim=0cm 0cm 0cm 0cm, clip=true, scale=0.35]{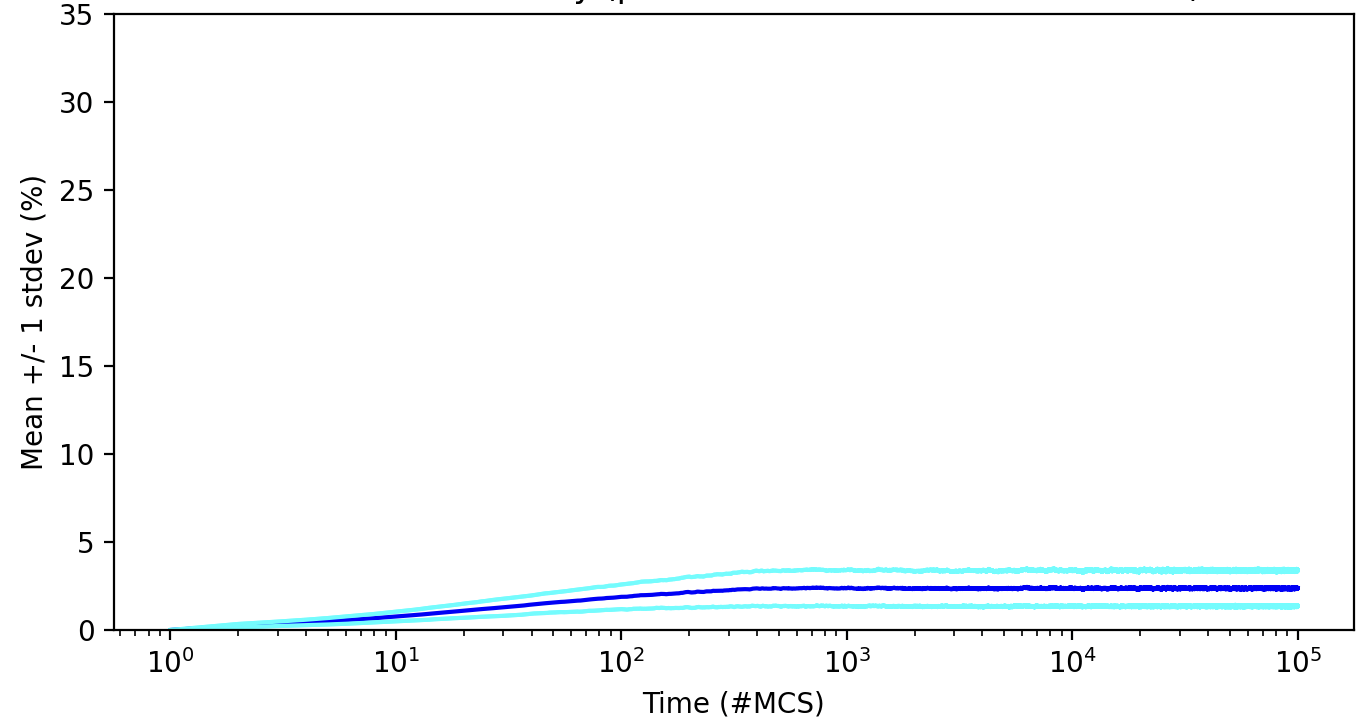}
\end{center}
\caption{Mean (plus and minus one standard deviation) of $\rho_V(t)$, the variation in species density at time $t$, over time, for the 100 IID unablated RES experiments ($N_a$$=$$0$) with  $L$$=$$150$ whose minimum densities were illustrated in Figure~\ref{fig:RES_NS05_NA0_L150_minD}; $\mu$$=$$\sigma$$=$$1.0;$ $M$$\in$$[ 10^{-7}, 10^{-3}]$. Format as for Figure~\ref{fig:NS05_NA1_L200_sdevs}.
}
\label{fig:RES_NS05_NA0_L150_stdevs}
\end{figure}

Figure~\ref{fig:RES_NS05_NA0_L150_minD}, shows that for RES, when the lattice side-length is increased to $L$$=$$150$, no underflow extinctions occur by $t$$=$200kMCS. Figure~\ref{fig:RES_NS05_NA0_L150_stdevs} shows the variation in densities $\rho_V(t)$ against time for the 100 IID $L$$=$$150$ RES experiments:  with no extinctions taking place, the RES $N_a$$=$$0$ system settles by $t$$\approx$$300$MCS to mean $\rho_V(t)$ of $\approx$$2.2$, roughly 80\% less than the $\rho_v(t)$ variations shown for the RES and OES $N_a$$=$$1$ experiments plotted in Figures~\ref{fig:NS05_NA1_L200_sdevs} and~\ref{fig:NS05_NA1_L200_Noop_sdevs}, respectively.

Having established that with RES at $L$$=$$150$ there is no evidence for species underflow extinctions at $N_a$$=$$0$, the extinctions seen in the $N_a$$=$$1$ RES results shown in Section~\ref{sec:resultsablated} (where $L$$=$$200$) can only have been caused by the ablations to the dominance network, rather being a consequence of running the experiments on too small a lattice. 

\subsubsection{Original Evolutionary Step (OES)}

\begin{figure}
\begin{center}
\includegraphics[trim=0cm 0cm 0cm 0cm, clip=true, scale=0.36]{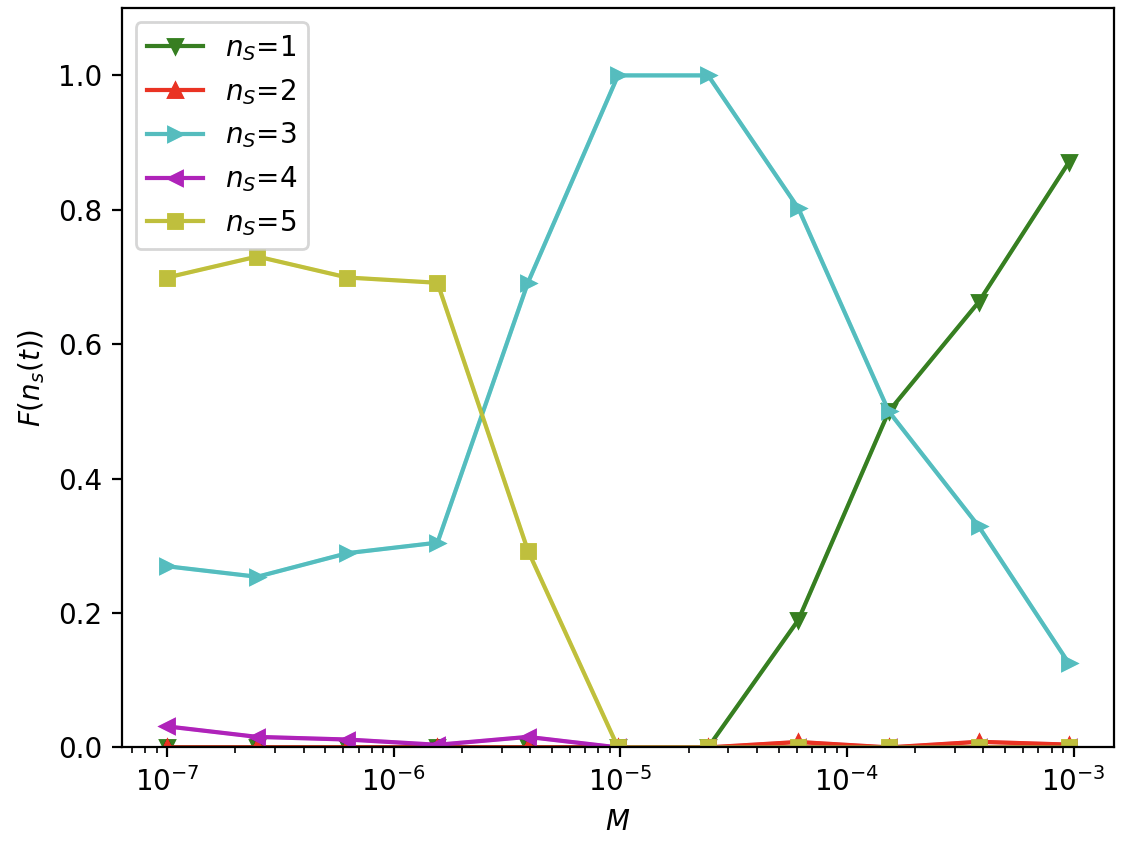}
\end{center}
\caption{Frequency distribution of species-counts at $t$$=$$100kMCS$ for OES experiments with no ablations ($N_a$$=$$0$): $L$$=$$200$; $\mu$$=$$\sigma$$=$$1.0$. At each value of $M$ sampled, 100 IID experiments were performed. 
Format as for Fig.~\ref{fig:Zhong_replicate_FvM_A}.
}
\label{fig:OES_NS05_NA0_L200_100k_freqs}
\end{figure}

Figure~\ref{fig:OES_NS05_NA0_L200_100k_freqs} shows the $F(n_s(t))$ outcome frequencies as $M$ is varied, resulting from unablated $N_a$$=$$0$ OES experiments with $L$$=$$200$ and  $\mu$$=$$\sigma$$=$$1.0$, at $t$$=$$100$kMCS which is the same duration of experiment studied by  \cite{zhong_etal_2022_ablatedRPSLS}. 

As can be seen, these outcomes are  similar to the $L$$=$$100$ results from RES that were illustrated in Figure~\ref{fig:RES_NS05_NA0_L100_FvM}: many of the simulations end with the number of surviving species being less than five, and as this is from the unablated dominance network the only plausible cause of the reduction in $n_s$ at the end of these experiments is underflow extinctions, and hence this is an indication that $L$$=$$200$ is simply too small for OES ESCGs. 
But -- crucially -- this is the value of $L$ that Zhong et al.\ used in their OES network ablation ($N_a$$>$$0$) experiments. 
Figure~\ref{fig:OES_NS05_NA0_L200_100k_freqs}  shows a sharp collapse in the frequency of $n_s(100\text{k})$$=$$5$ outcomes occurring over the range $M$$\in$$[10^{-6},10^{-5}]$, with all simulations showing $n_s(100\text{k})$$=$$3$ at $M$$=$$10^{-5}$, but the $n_s$$=$$3$ system is clearly not stable because $F(n_s(100\text{k})$$=$$3)$ then falls steadily as $M$ increases over $M$$=$$[2$$\times$$10^{-5}, 2$$\times$$10^{-3}]$, being replaced by a consequential rise in $F(n_s(100\text{k})$$=$$1)$.  

\begin{figure}
\begin{center}
\includegraphics[trim=0cm 0cm 0cm 0cm, clip=true, scale=0.36]{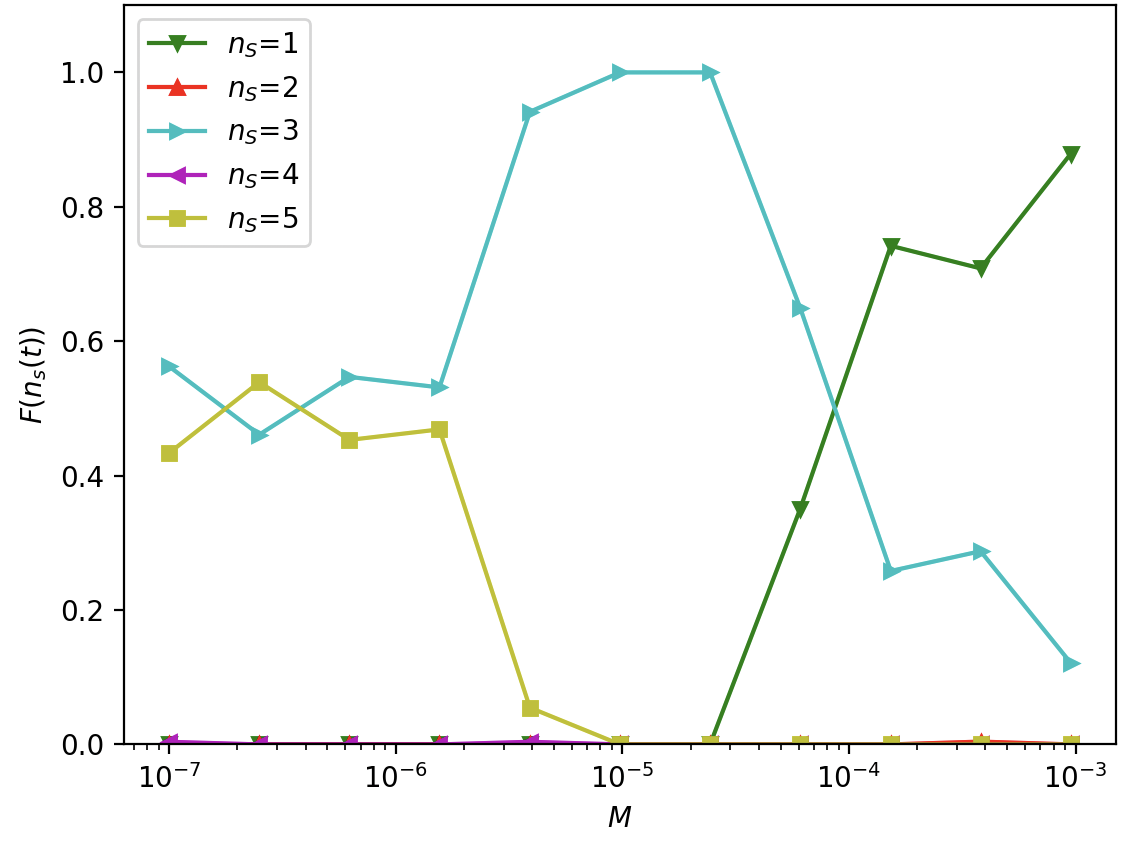}
\end{center}
\caption{Frequency distribution of species-counts at $t$$=$$200$kMCS 
(twice as long as in Figure~\ref{fig:OES_NS05_NA0_L200_100k_freqs}) 
for OES experiments with no ablations ($N_a$$=$$0$): $L$$=$$200$; $\mu$$=$$\sigma$$=$$1.0$. At each value of $M$ sampled, 100 IID experiments were performed. 
Format as for Fig.~\ref{fig:Zhong_replicate_FvM_A}.
}
\label{fig:OES_NS05_NA0_L200_200k_freqs}
\end{figure}

Furthermore, as is demonstrated in Figure~\ref{fig:OES_NS05_NA0_L200_200k_freqs}, the results at $t$$=$$100$kMCS are just a snapshot of an evolving system that is still far from any asymptotic state. Figure~\ref{fig:OES_NS05_NA0_L200_200k_freqs} shows outcomes from the same simulations as were visualized at $t$$=$$100$kMCS in Figure~\ref{fig:OES_NS05_NA0_L200_100k_freqs}, but run for twice as long, out to $t$$=$$200$kMCS. In these longer-duration experiments we see the frequency of $n_s$$=$$5$ outcomes for low $M$ has dropped from roughly 70\% at $t$$=$$100$kMCS to roughly 45\% at $t$$=$$200$kMCS. Presumably if we ran these experiments for a lot longer, out to $t$$=$$1000$kMCS for example, we would see no  $n_s$$=$$5$ outcomes and only $n_s$$=$$3$ for low $M$, and then, around $M$$=$$10^{-4}$, a sharp collapse in $n_s$$=$$3$, being replaced by $n_s$$=$$1$. 

That is, at $L$$=$$200$, the relationship between the true long-term asymptotic number of coexisting species and agent mobility $M$ in the {\em unablated}\/ OES system can be described in exactly the same words as that relationship in the $N_a$$=$$1$ {\em ablated} OES system: for low values of $M$, the asymptotic state is $F(n_s(t)$$=$$3)$$=$$1.0$ and then as $M$ is increased through the range $M$$\in$$[10^{-5},10^{-4}]$ the frequency  $F(n_s(t)$$=$$3)$ collapses to zero and instead  $F(n_s(t)$$\leq$$2)$ rises sharply to 1.0. Apparently the primary effect of a single ablation is that you don't have to wait so long, don't have to run so many MCSs, before you see the collapse occur.

\begin{figure}
\begin{center}
\includegraphics[trim=0cm 0.0cm 0cm 0cm, clip=true, scale=0.35]{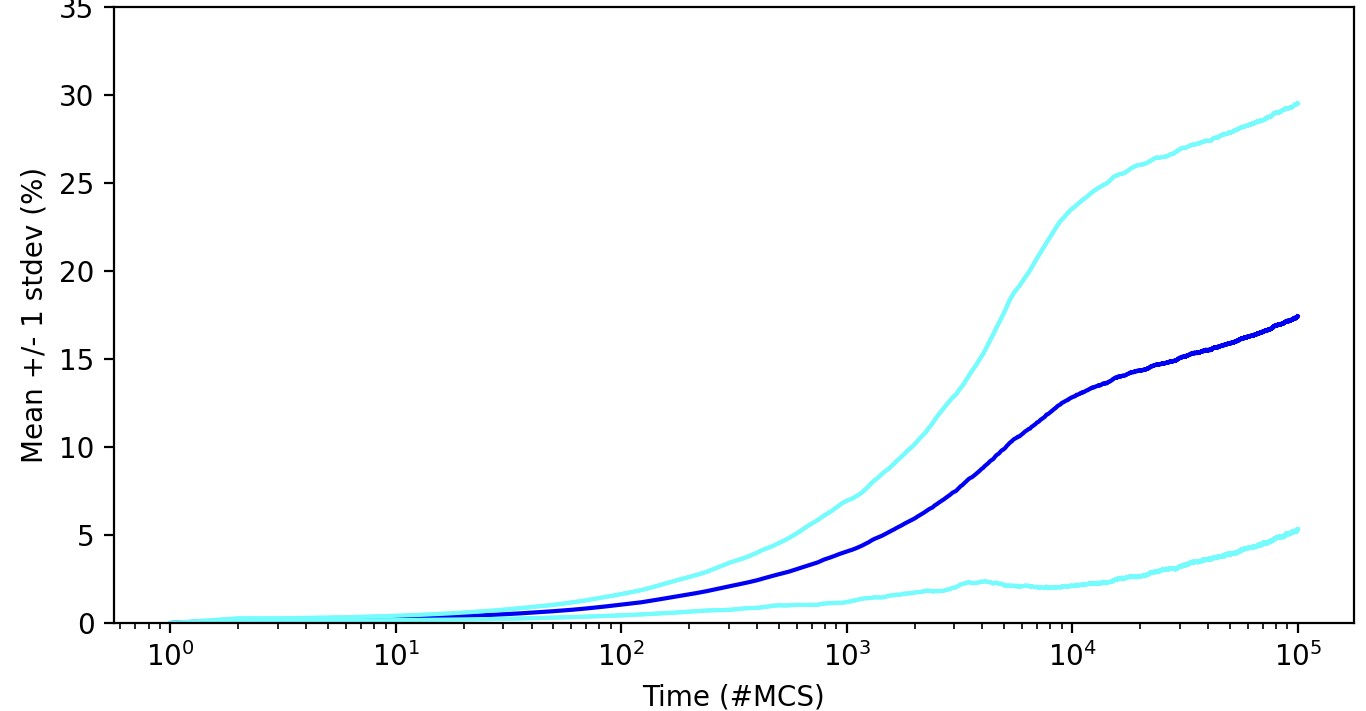}
\end{center}
\caption{Mean (plus and minus one standard deviation) $\rho_V(t)$  over time, for unablated OES experiments ($N_a$$=$$0$) with  $L$$=$$200$; $\mu$$=$$\sigma$$=$$1.0;$ $M \in [ 10^{-7}, 10^{-3}]$. Format as for Figure~\ref{fig:NS05_NA1_L200_sdevs}. There is no evidence here of the system approaching an asymptote before or at  $t$$=$$10^5$, despite this being the duration of simulations run to identify the ``asymptotic state'' in \cite{zhong_etal_2022_ablatedRPSLS}.
}
\label{fig:OES_NS05_NA0_L200_stdevs}
\end{figure}

Figure~\ref{fig:OES_NS05_NA0_L200_stdevs} shows the variation in species density for the unablated ($N_a$$=$$0$) OES system, a stark contrast to the comparable plot for the RES system of Figure~\ref{fig:RES_NS05_NA0_L150_stdevs}.
The revelation here that the $L$$=$$200$ OES system is too small to avoid species underflow extinctions casts some doubt on the results published in \cite{zhong_etal_2022_ablatedRPSLS}: they offer no discussion of the possibility that their choice of $L$ is too small for the expected variability in the population dynamics.

Exploratory experiments, repeating the 200kMCS $N_a$$=$$0$ OES simulations over the same range of $M$ values but with $L$$=$$400$ (which increases $N$ by 300\%, relative to the $L$$=$$200$ used by Zhong et al.), showed stable evolution (i.e., five species as the asymptotic state) for values of $M$$<$$10^{-6}$ but species underflow extinctions occur once mobility $M$ is increased beyond that threshold. It seems reasonable to expect that at some higher value of $L$, the OES system will be stable across all values of $M$ of interest, but even the $L$$=$$400$ result tells us enough: five-species ESCGs can be stably simulated using RES at $L$$=$$150$, but attempting to run evolutionarily stable simulations using OES will require at least $L$$=$$400$. Because the run-times of these simulations scales with $N$ (the square of $L$), and because $150^2$$/$$400^2$$\approx$$0.14$, the results presented in this paper indicate that, for the ESCGs studied here, stable evolutionary simulations using RES will require no more than 14\% of the run-time needed for the same simulation to execute when using OES: a reduction of at least 85\%.

\section{Discussion}
\label{sec:discussion}

The fact that the RES experiments for $N_a$$=$$0$ are stable (no species underflow extinctions) at $L$$=$$150$ but the corresponding OES experiments remain unstable at  $L$$=$$400$ is explainable by reference to the observation made in discussing Figures~\ref{fig:NS05_NA1_L200_sdevs} and~\ref{fig:NS05_NA1_L200_Noop_sdevs}: the variance in species densities is markedly higher in the OES system than in the RES system. And the more variability there is in species density in an ESCG, the bigger the lattice needs to be to avoid underflow extinctions within the duration of the experiment.

\section{Conclusion}
\label{sec:conclusion}
The novel contributions of this paper have been my introduction of the Revised Elementary Step (RES) as a replacement for the Original Elementary Step (OES) in this class of Evolutionary Spatial Cyclic Games (ESCGs). The introduction of RES was motivated by the desire to waste less time in running ESCG simulations that perform large numbers of no-ops, and in which large numbers of the lattice's cells can be empty, further wasting compute cycles. However the RES's expanded use of the random-number generator means that the observable reductions in like-for-like run-times were only 10\% or so. Nevertheless, an unexpected effect of the changes introduced in RES is that the co-evolutionary population dynamics of the RPSLS model when using RES are much less volatile than when OES is used instead, and this facilitates the study of interesting research questions using much smaller lattices in RES-based simulation studies than in OES-based ones. I have shown here that the smaller total cell-counts needed for stable evolution in RES-based systems give reductions in simulation run-times of 85\% or more, a big saving.



\bibliography{../../dc_bibliography}

\begin{thebibliography}{}

\bibitem[\protect\astroncite{Avelino
  et~al.}{2022}]{avelino_deoliveria_trintin_2022_RPS_bigN}
Avelino, P., {de Oliveria}, B., and Trintin, R. (2022).
\newblock Parity effects in rock-paper-scissors type models with a number of
  species ns $\leq$ 12.
\newblock {\em Chaos, Solitons and Fractals}, 155(111738).

\bibitem[\protect\astroncite{Bazeia
  et~al.}{2022}]{bazeia_bongestab_deoliveira_2022_RPS}
Bazeia, D., Bongestab, M., and {de Oliveira}, B. (2022).
\newblock Influence of the neighborhood on cyclic models of biodiversity.
\newblock {\em Physica A}, 587(126547).

\bibitem[\protect\astroncite{Cheng
  et~al.}{2014}]{cheng_yao_huang_park_do_lai_2014}
Cheng, H., Yao, Z., Huang, Z.-G., Park, J., Do, Y., and Lai, Y.-C. (2014).
\newblock Mesoscopic interactions and species coexistence in evolutionary game
  dynamics of cyclic competitions.
\newblock {\em Nature Scientific Reports}, 4(7486).

\bibitem[\protect\astroncite{Cliff}{2024}]{cliff_2024_circulants}
Cliff, D. (2024).
\newblock {Tournament versus Circulant: On Simulating 7-Species Evolutionary
  Spatial Cyclic Games with Ablated Predator-Prey Networks as Models of
  Biodiversity}.
\newblock In {\em {Proceedings of the 36th European Modelling and Simulation
  Symposium (EMSS2024)}}.

\bibitem[\protect\astroncite{Kabir and
  Tanimoto}{2021}]{kabir_tanimoto_2021_RPS}
Kabir, K. and Tanimoto, J. (2021).
\newblock The role of pairwise nonlinear evolutionary dynamics in the
  rock–paper–scissors game with noise.
\newblock {\em Applied Mathematics and Computation}, 394(125767).

\bibitem[\protect\astroncite{Kass and Bryla}{1998}]{kass_bryla_1998}
Kass, S. and Bryla, K. (1998).
\newblock {Rock Paper Scissors Spock Lizard}.
\newblock \url{samkass.com/theories/RPSSL.html}.

\bibitem[\protect\astroncite{Kubyana et~al.}{2024}]{kubyana_landi_hui_2024_RPS}
Kubyana, M., Landi, P., and Hui, C. (2024).
\newblock Adaptive rock-paper-scissors game enhances eco-evolutionary
  performance at cost of dynamic stability.
\newblock {\em Applied Mathematics and Computation}, 468(128535).

\bibitem[\protect\astroncite{Laird and Schamp}{2006}]{laird_schamp_2006_RPSLS}
Laird, R. and Schamp, B. (2006).
\newblock Competitive intransitivity promotes species coexistence.
\newblock {\em American Naturalist}, 168:182--193.

\bibitem[\protect\astroncite{Laird and Schamp}{2008}]{laird_schamp_2008_RPSLS}
Laird, R. and Schamp, B. (2008).
\newblock Does local competition increase the coexistence of species in
  intransitive networks.
\newblock {\em Ecology}, 89:237--247.

\bibitem[\protect\astroncite{Laird and
  Schamp}{2009}]{laird_schamp_2009_coexistence_RPSLS}
Laird, R. and Schamp, B. (2009).
\newblock Species coexistence, intransitivity, and topological variation in
  competitive tournaments.
\newblock {\em J.\ Theoretical Biology}, 256:90--95.

\bibitem[\protect\astroncite{May and Leonard}{1975}]{may_leonard_1975}
May, R. and Leonard, W. (1975).
\newblock Nonlinear aspects of competition between species.
\newblock {\em {SIAM Journal of Applied Mathematics}}, 29:243--253.

\bibitem[\protect\astroncite{Menezes
  et~al.}{2022a}]{menezes_batista_rangel_2022_RPS}
Menezes, J., Batista, S., and Rangel, E. (2022a).
\newblock Spatial organisation plasticity reduces disease infection risk in
  rock–paper–scissors models.
\newblock {\em Biosystems}, 221(104777).

\bibitem[\protect\astroncite{Menezes
  et~al.}{2022b}]{menezes_rangel_moura_2022_RPS}
Menezes, J., Rangel, E., and Moura, B. (2022b).
\newblock Aggregation as an antipredator strategy in the rock-paper-scissors
  model.
\newblock {\em Ecological Informatics}, 69(101606).

\bibitem[\protect\astroncite{Menezes et~al.}{2023}]{menezes_barbalho_2023_RPS}
Menezes, J., Rodrigues, S., and Batista, S. (2023).
\newblock Mobility unevenness in rock–paper–scissors models.
\newblock {\em Ecological Complexity}, 52(101028).

\bibitem[\protect\astroncite{Mood and Park}{2021}]{mohd_park_2021_RPS}
Mood, M. and Park, J. (2021).
\newblock The interplay of rock-paper-scissors competition and environments
  mediates species coexistence and intriguing dynamics.
\newblock {\em Chaos, Solitons and Fractals}, 153(111579).

\bibitem[\protect\astroncite{Nagatani
  et~al.}{2018}]{nagatani_ichinose_tainaka_2018_RPS}
Nagatani, T., Ichinose, G., and Tainaka, K. (2018).
\newblock Metapopulation model for rock–paper–scissors game: Mutation
  affects paradoxical impacts.
\newblock {\em Journal of Theoretical Biology}, 450(22--29).

\bibitem[\protect\astroncite{Park}{2021}]{park_2021_RPS}
Park, J. (2021).
\newblock Evolutionary dynamics in the rock-paper-scissors system by changing
  community paradigm with population flow.
\newblock {\em Chaos, Solitons and Fractals}, 142(110424).

\bibitem[\protect\astroncite{Park and Jang}{2019}]{park_jang_2019}
Park, J. and Jang, B. (2019).
\newblock Robust coexistence with alternative competition strategy in the
  spatial cyclic game of five species.
\newblock {\em Chaos}, 29(051105).

\bibitem[\protect\astroncite{Park and Jang}{2023}]{park_jang_2023_RPS}
Park, J. and Jang, B. (2023).
\newblock Role of adaptive intraspecific competition on collective behavior in
  the rock–paper–scissors game.
\newblock {\em Chaos, Solitons and Fractals}, 171(113448).

\bibitem[\protect\astroncite{Reichenbach
  et~al.}{2007a}]{reichenbach_mobilia_frey_2007_nature}
Reichenbach, T., Mobilia, M., and Frey, E. (2007a).
\newblock {Mobility promotes and jeopardizes biodiversity in
  rock-paper-scissors games}.
\newblock {\em Nature}, 448(06095).

\bibitem[\protect\astroncite{Reichenbach
  et~al.}{2007b}]{reichenbach_mobilia_frey_2007_physrevlet}
Reichenbach, T., Mobilia, M., and Frey, E. (2007b).
\newblock {Noise and Correlations in a Spatial Population Model with Cyclic
  Competition}.
\newblock {\em Physical Review Letters}, 99(238105).

\bibitem[\protect\astroncite{Reichenbach
  et~al.}{2008}]{reichenbach_mobilia_frey_2008_jtb}
Reichenbach, T., Mobilia, M., and Frey, E. (2008).
\newblock {Self-Organization of Mobile Populations in Cyclic Competition}.
\newblock {\em Journal of Theoretical Biology}, 254(2008):363--383.

\bibitem[\protect\astroncite{Viswanathan et~al.}{2024}]{viswanathan_etal_2024}
Viswanathan, K., Wilson, A., Bhattacharyya, S., and Hens, C. (2024).
\newblock Ecological resilience in a circular world: Mutation and extinction in
  five-species ecosystems.
\newblock {\em Chaos, Solitons, and Fractals}, 180(114548).

\bibitem[\protect\astroncite{Wolfram}{2002}]{wolfram_2002_book}
Wolfram, S. (2002).
\newblock {\em {A New Kind of Science}}.
\newblock Wolfram Media.

\bibitem[\protect\astroncite{Zhang
  et~al.}{2022}]{zhang_bearup_guo_zhang_liao_2022_RPS}
Zhang, Z., Bearup, D., Guo, G., Zhang, H., and Liao, J. (2022).
\newblock Competition modes determine ecosystem stability in
  rock–paper–scissors games.
\newblock {\em Physica A}, 607(128176).

\bibitem[\protect\astroncite{Zhong et~al.}{2022}]{zhong_etal_2022_ablatedRPSLS}
Zhong, L., Zhang, L., Li, H., Dai, Q., and Yang, J. (2022).
\newblock Species coexistence in spatial cyclic game of five species.
\newblock {\em Chaos, Solitons, and Fractals}, 156(111806).

\end{thebibliography}

\clearpage
\newpage


\section*{Appendix A: Original ES (OES), $L$$=$$200$, $N_a$$=$$0$}

\begin{figure}[h]
\begin{center}
\includegraphics[trim=0cm 0cm 0cm 0cm, clip=true,scale=0.4]{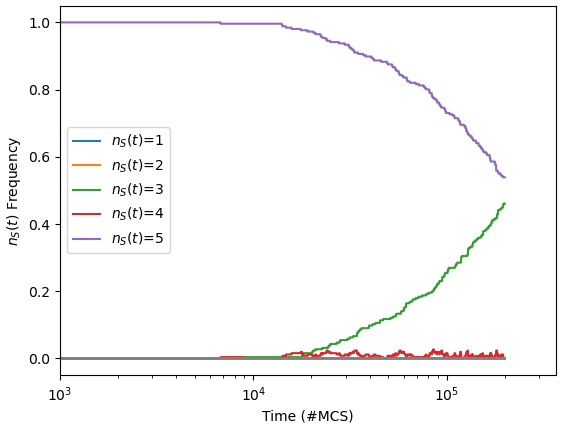}
\includegraphics[trim=0cm 0cm 0cm 0cm, clip=true,scale=0.4]{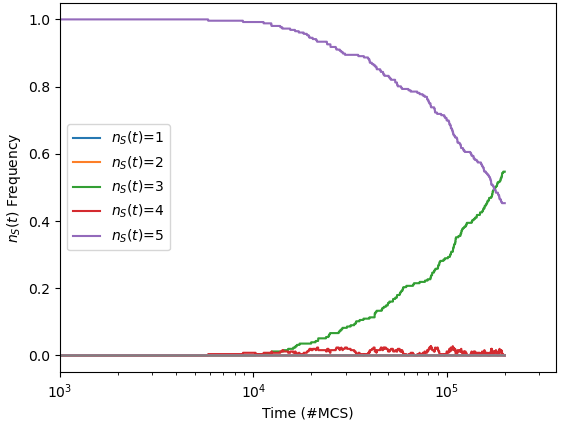}
\end{center}
\caption{Frequency distribution of species-counts at $t$$=$$200$kMCS for OES experiments with no ablations ($N_a$$=$$0$): $L$$=$$200$; $\mu$$=$$\sigma$$=$$1.0$; 
$M$$\in$$\{2.50$$\times$$10^{-7}, 6.25$$\times$$10^{-7}\}$ (from top to bottom). 
At each value of $M$ sampled, 250 IID experiments were performed. Horizontal axis is time, measured in MCS; vertical axis is what proportion of the IID simulations had $n_S(t)$$=$$c$ at each time $t$ for $c$$\in$$\{1, \ldots, N_S\}$.
}
\label{fig:OES_NS05_NA0_L200_A}
\end{figure}

\begin{figure}
\begin{center}
\includegraphics[trim=0cm 0cm 0cm 0cm, clip=true,scale=0.4]{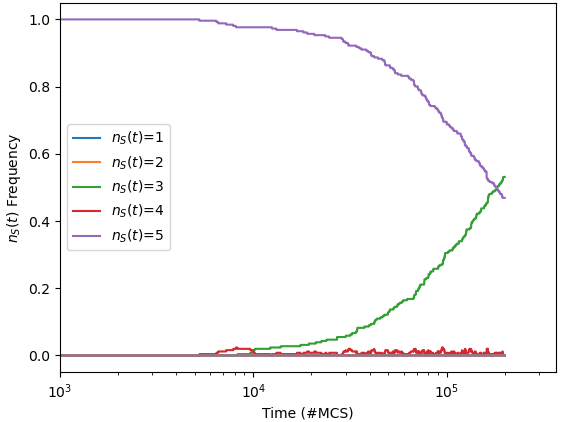}
\includegraphics[trim=0cm 0cm 0cm 0cm, clip=true,scale=0.4]{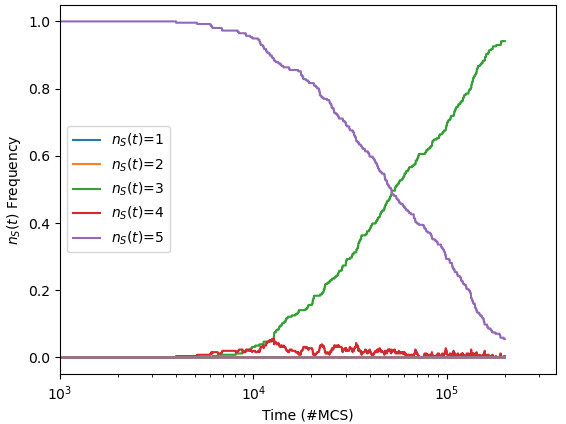}
\includegraphics[trim=0cm 0cm 0cm 0cm, clip=true,scale=0.4]{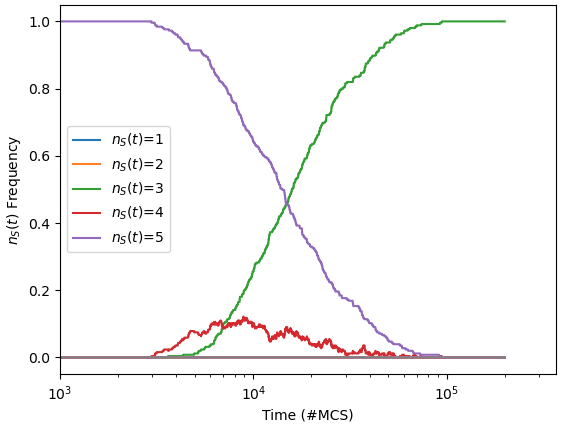}
\end{center}
\caption{Frequency distribution of species-counts at $t$$=$$200$kMCS  for OES experiments with no ablations ($N_a$$=$$0$): $L$$=$$200$; $\mu$$=$$\sigma$$=$$1.0$; 
$M$$\in$$\{1.56$$\times$$10^{-6}, 3.91$$\times$$10^{-6}, 9.77$$\times$$10^{-6} \}$ (from top to bottom). 
At each value of $M$ sampled, 250 IID experiments were performed. Format as for Fig.~\ref{fig:OES_NS05_NA0_L200_A}.
}
\label{fig:OES_NS05_NA0_L200_B}
\end{figure}

\begin{figure}
\begin{center}
\includegraphics[trim=0cm 0cm 0cm 0cm, clip=true,scale=0.4]{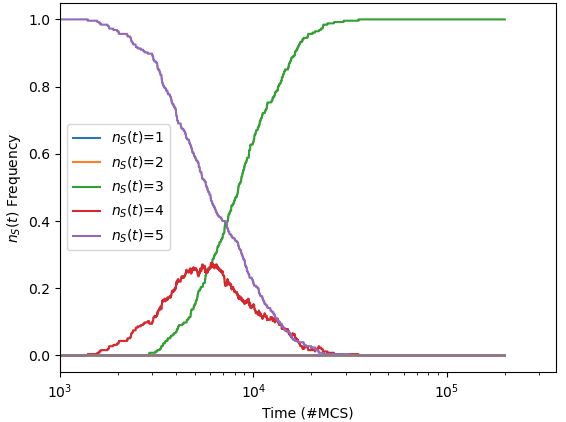}
\includegraphics[trim=0cm 0cm 0cm 0cm, clip=true,scale=0.4]{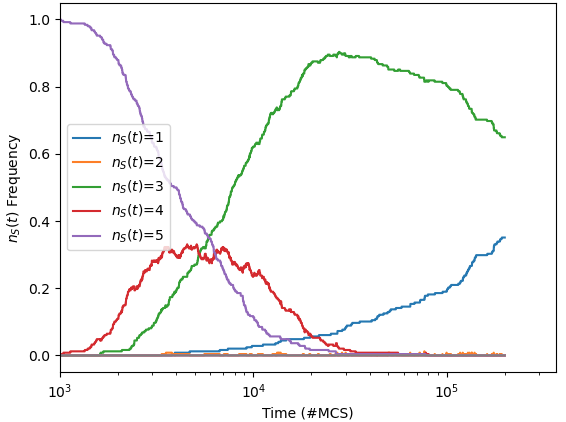}
\includegraphics[trim=0cm 0cm 0cm 0cm, clip=true,scale=0.4]{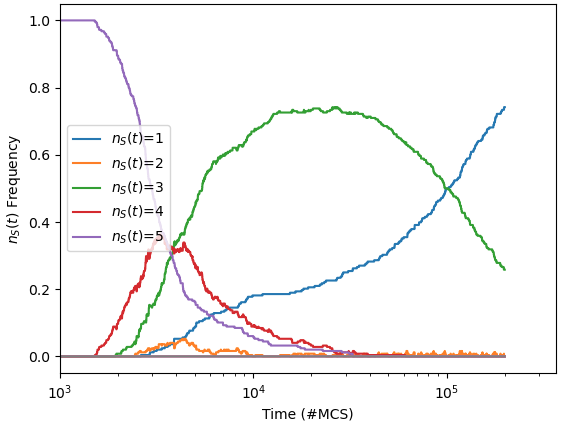}
\end{center}
\caption{Frequency distribution of species-counts at $t$$=$$200$kMCS  for OES experiments with no ablations ($N_a$$=$$0$): $L$$=$$200$; $\mu$$=$$\sigma$$=$$1.0$; 
$M$$\in$$\{2.44$$\times$$10^{-5}, 6.10$$\times$$10^{-5}, 1.53$$\times$$10^{-4} \}$ (from top to bottom). 
At each value of $M$ sampled, 250 IID experiments were performed. Format as for Fig.~\ref{fig:OES_NS05_NA0_L200_A}.
}
\label{fig:OES_NS05_NA0_L200_C}
\end{figure}

Figures~\ref{fig:OES_NS05_NA0_L200_A} to~\ref{fig:OES_NS05_NA0_L200_C} show time series of the evolution the number of surviving species $n_S(t)$ over 200kMCS for the OES ESCG with no ablations ($N_a$$=$$0$), $L$$=$$200$ and $\mu$$=$$\sigma$$=$$1.0$ for various values of $M$$\in$$[ 10^{-7}, 10^{-3}]$, with 250 IID simulations executed for each value of $M$ sampled. 

Given that $N_a$$=$$0$, the progressive reductions in $n_s(t)$ seen in all these time-series are each underflow extinctions, i.e.\ they are a consequence of the variation in population dynamics inherent in the five-species RPSLS model running up against the discretization limit of $L$$=$$200$.

\begin{figure}
\begin{center}
\includegraphics[trim=0cm 0cm 0cm 0cm, clip=true,scale=0.4]{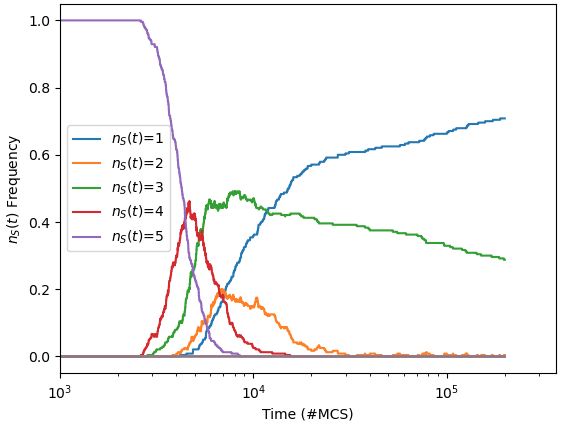}
\includegraphics[trim=0cm 0cm 0cm 0cm, clip=true,scale=0.4]{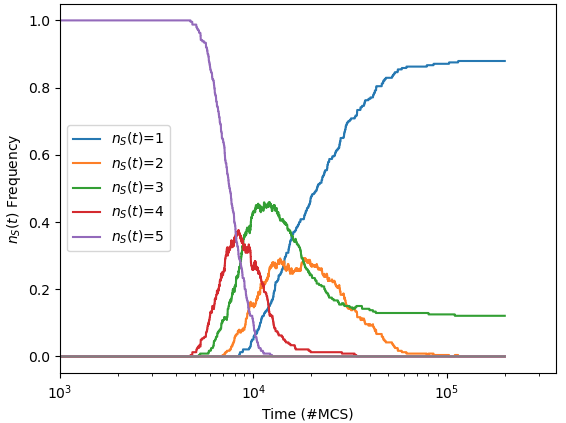}
\end{center}
\caption{Frequency distribution of species-counts at $t$$=$$200$kMCS  for OES experiments with no ablations ($N_a$$=$$0$): $L$$=$$200$; $\mu$$=$$\sigma$$=$$1.0$; 
$M$$\in$$\{ 3.81$$\times$$10^{-4}, 9.54$$\times$$10^{-4} \}$ (from top to bottom). 
At each value of $M$ sampled, 250 IID experiments were performed. Format as for Fig.~\ref{fig:OES_NS05_NA0_L200_A}.
}
\label{fig:OES_NS05_NA0_L200_D}
\end{figure}

\clearpage

\section*{Appendix B: Original ES (OES), $L$$=$$200$, $N_a$$=$$1$}

\begin{figure}[h]
\begin{center}
\includegraphics[trim=0cm 0cm 0cm 0cm, clip=true,scale=0.4]{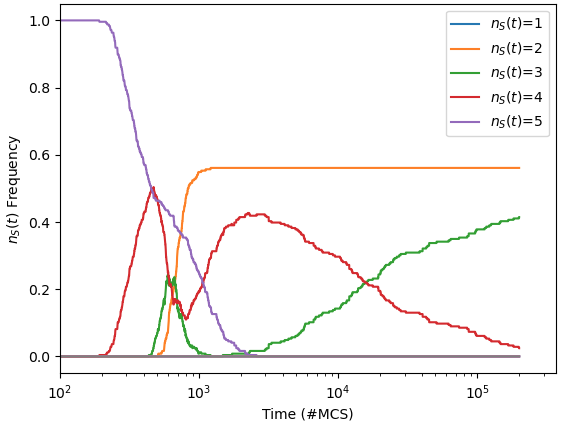}
\end{center}
\caption{Frequency distribution of species-counts to $t$$=$$200$kMCS  for OES experiments with one ablation ($N_a$$=$$1$): $L$$=$$200$; $\mu$$=$$\sigma$$=$$1.0$; $M$$\i=$$10^{-7}$. 
Results from 250 IID simulation runs.
Format as for Fig.~\ref{fig:OES_NS05_NA0_L200_A}.
}
\label{fig:OES_NS05_NA1_L200_A}
\end{figure}

\begin{figure}
\begin{center}
\includegraphics[trim=0cm 0cm 0cm 0cm, clip=true,scale=0.4]{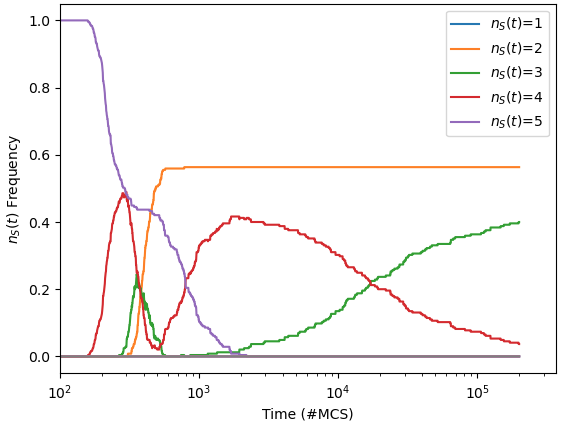}
\includegraphics[trim=0cm 0cm 0cm 0cm, clip=true,scale=0.4]{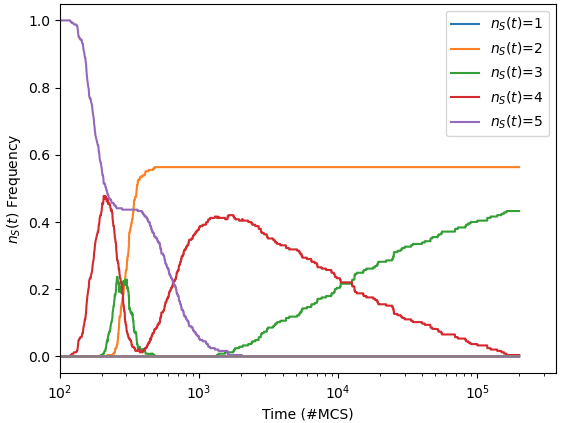}
\includegraphics[trim=0cm 0cm 0cm 0cm, clip=true,scale=0.4]{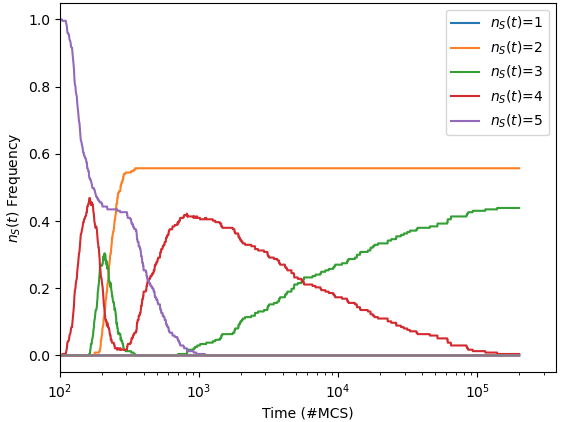}
\end{center}
\caption{Frequency distribution of species-counts to $t$$=$$200$kMCS  for OES experiments with one ablation ($N_a$$=$$1$): $L$$=$$200$; $\mu$$=$$\sigma$$=$$1.0$; 
$\mu$$=$$\sigma$$=$$1.0$; 
$M$$\in$$\{6.25$$\times$$10^{-7}, 1.56$$\times$$10^{-6}, 3.91$$\times$$10^{-6}\}$ (top to bottom). 
At each value of $M$ sampled, 250 IID experiments were performed. Format as for 
Fig.~\ref{fig:OES_NS05_NA0_L200_A}.
}
\label{fig:OES_NS05_NA1_L200_B}
\end{figure}

\begin{figure}
\begin{center}
\includegraphics[trim=0cm 0cm 0cm 0cm, clip=true,scale=0.4]{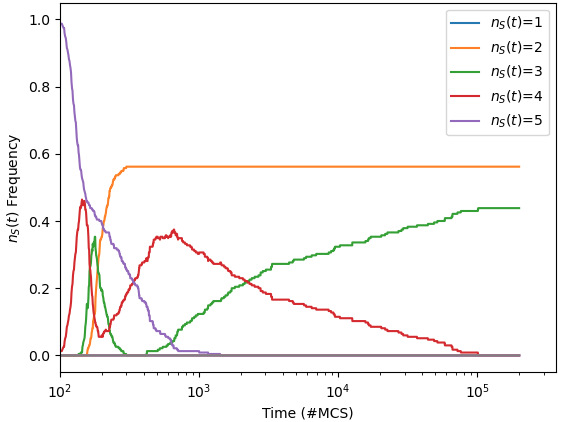}
\includegraphics[trim=0cm 0cm 0cm 0cm, clip=true,scale=0.4]{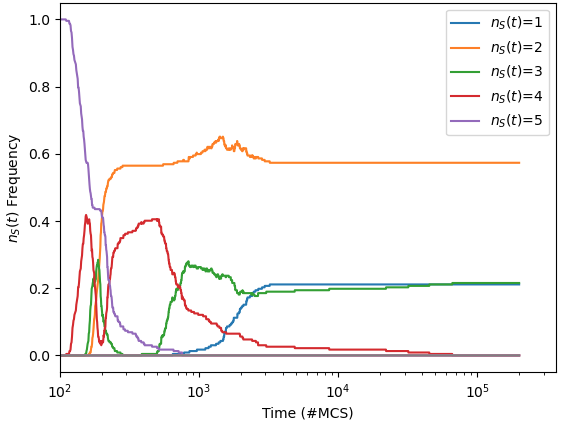}
\includegraphics[trim=0cm 0cm 0cm 0cm, clip=true,scale=0.4]{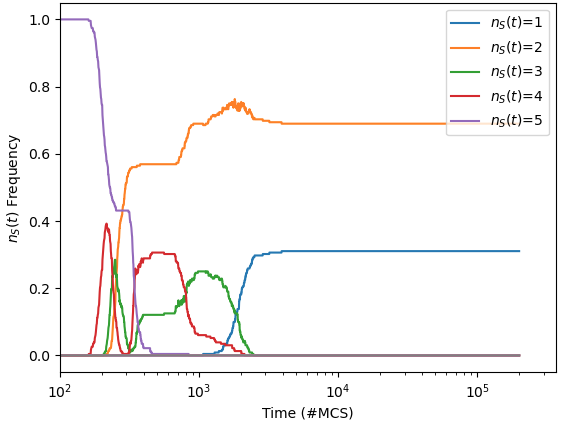}
\end{center}
\caption{Frequency distribution of species-counts to $t$$=$$200$kMCS  for OES experiments with one ablation ($N_a$$=$$1$): $L$$=$$200$; $\mu$$=$$\sigma$$=$$1.0$;
$\mu$$=$$\sigma$$=$$1.0$; 
$M$$\in$$\{9.77$$\times$$10^{-6}, 2.44$$\times$$10^{-5}, 6.10$$\times$$10^{-5}\}$ (top to bottom). 
At each value of $M$ sampled, 250 IID experiments were performed. Format as for 
Fig.~\ref{fig:OES_NS05_NA0_L200_A}.
}
\label{fig:OES_NS05_NA1_L200_C}
\end{figure}

\begin{figure}
\begin{center}
\includegraphics[trim=0cm 0cm 0cm 0cm, clip=true,scale=0.4]{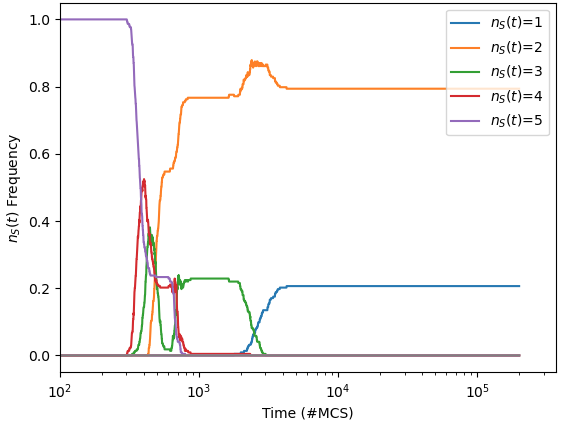}
\includegraphics[trim=0cm 0cm 0cm 0cm, clip=true,scale=0.4]{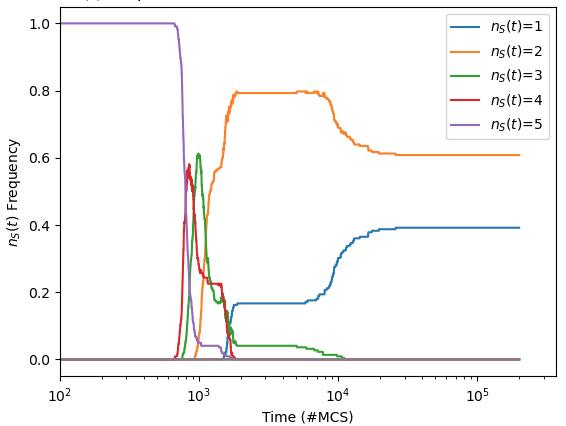}
\includegraphics[trim=0cm 0cm 0cm 0cm, clip=true,scale=0.4]{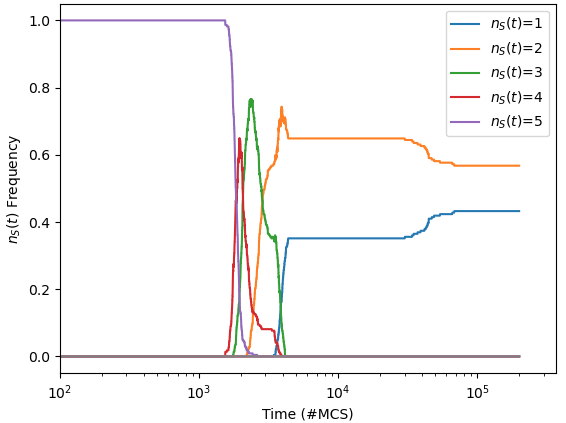}
\end{center}
\caption{Frequency distribution of species-counts to $t$$=$$200$kMCS  for OES experiments with one ablation ($N_a$$=$$1$): $L$$=$$200$; $\mu$$=$$\sigma$$=$$1.0$;
$\mu$$=$$\sigma$$=$$1.0$; $M$$\in$$\{1.53$$\times$$10^{-4}, 3.81$$\times$$10^{-4}, 9.54$$\times$$10^{-4}\}$ (top to bottom). 
At each value of $M$ sampled, 250 IID experiments were performed. Format as for 
Fig.~\ref{fig:OES_NS05_NA0_L200_A}.
}
\label{fig:OES_NS05_NA1_L200_D}
\end{figure}

Figures~\ref{fig:OES_NS05_NA1_L200_A} to~\ref{fig:OES_NS05_NA1_L200_D} show time series of the evolution the number of surviving species $n_S(t)$ over 200kMCS for the OES ESCG with one ablation (i.e., ablating either the $S_0$$\rightarrow$$S_1$ or the $S_0$$\rightarrow$$S_3$ directed edge from the dominance network, $N_a$$=$$1$), $L$$=$$200$ and $\mu$$=$$\sigma$$=$$1.0$ for various values of $M$$\in$$[ 10^{-7}, 10^{-3}]$, with 250 IID simulations executed for each value of $M$ sampled. 

As can be seen in Figure~\ref{fig:OES_NS05_NA1_L200_A}, even when given twice as many MCS as were used by Zhong et al.\ to determine the ``asymptotic'' behavior of the ESCG, the system has still not fully settled to a steady state because at $M$$=$$10^{-7}$ the frequency of $n_s(t)$$=$$4$ is falling steadily but has not yet reached zero: from informally ``eyeballing'' the graphs in Figure~\ref{fig:OES_NS05_NA1_L200_A}, it seems that for low $M$ the system would need to be simulated out to $t$$=$$300$kMCS or more before it reaches a steady state.

Furthermore, Figures~\ref{fig:OES_NS05_NA1_L200_B} and~\ref{fig:OES_NS05_NA1_L200_C} reveal that, at higher values of $M$, the system can still be part-way through a transient at 200kMCS, with the frequency of $n_S(t)$$=$$3$ falling steadily, and  $n_S(t)$$=$$1$ rising in consequence, and again it seems that the system is unlikely to converge on a true asymptotic state until several hundred thousand more MCS are simulated.

\clearpage

\section*{Appendix C: Revised ES (RES), $L$$=$$200$, $N_a$$=$$1$}

\begin{figure}[h]
\begin{center}
\includegraphics[trim=0cm 0cm 0cm 0cm, clip=true,scale=0.4]{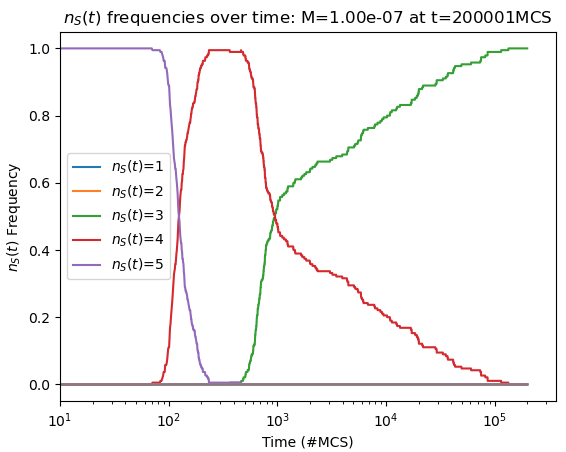}
\end{center}
\caption{Frequency distribution of species-counts to $t$$=$$200$kMCS  for RES experiments with one ablation ($N_a$$=$$1$): $L$$=$$200$; $\mu$$=$$\sigma$$=$$1.0$; $M$$=$$10^{-7}$; results computed over outcome from 175 IID simulation runs.  
Format as for Fig.~\ref{fig:OES_NS05_NA0_L200_A}.
}
\label{fig:RES_NS05_NA1_L200_A}
\end{figure}

\begin{figure}
\begin{center}
\includegraphics[trim=0cm 0cm 0cm 0cm, clip=true,scale=0.4]{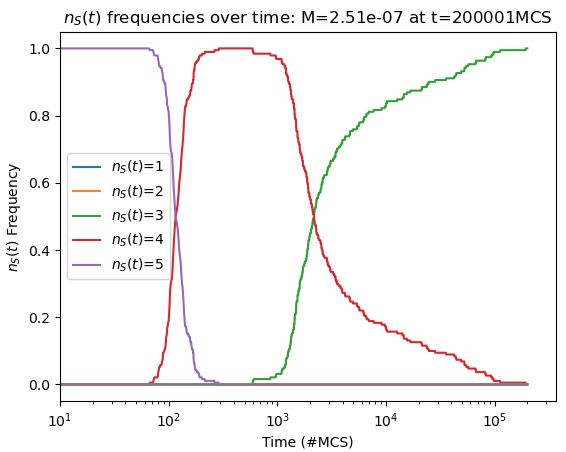}
\includegraphics[trim=0cm 0cm 0cm 0cm, clip=true,scale=0.4]{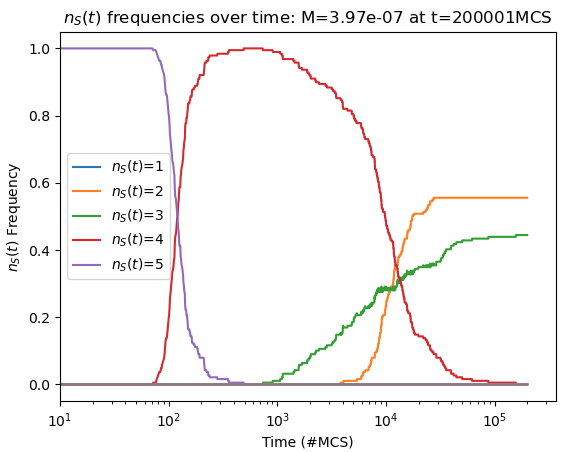}
\includegraphics[trim=0cm 0cm 0cm 0cm, clip=true,scale=0.4]{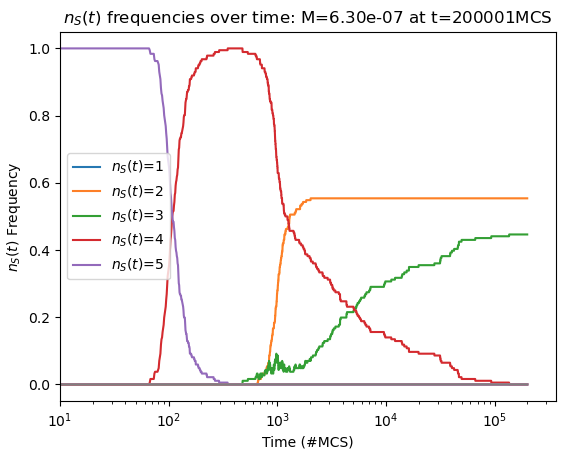}
\end{center}
\caption{Frequency distribution of species-counts to $t$$=$$200$kMCS  for RES experiments with one ablation ($N_a$$=$$1$): $L$$=$$200$; $\mu$$=$$\sigma$$=$$1.0$; 
$\mu$$=$$\sigma$$=$$1.0$; 
$M$$\in$$\{2.51$$\times$$10^{-7}, 3.97$$\times$$10^{-7}, 6.30$$\times$$10^{-7}\}$ (from top to bottom). 
At each value of $M$ sampled, 175 IID experiments were performed. 
Format as for  Fig.~\ref{fig:OES_NS05_NA0_L200_A}.
}
\label{fig:RES_NS05_NA1_L200_B}
\end{figure}

\begin{figure}
\begin{center}
\includegraphics[trim=0cm 0cm 0cm 0cm, clip=true,scale=0.4]{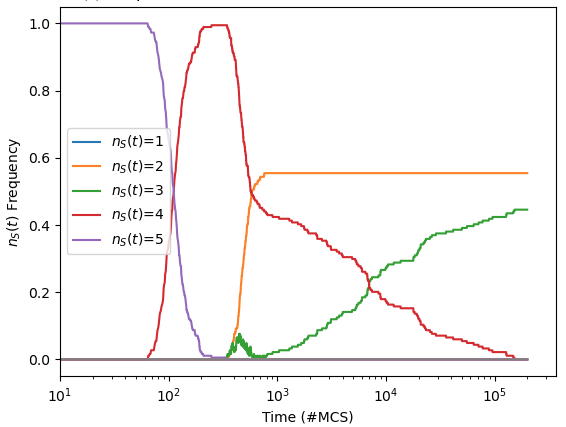}
\includegraphics[trim=0cm 0cm 0cm 0cm, clip=true,scale=0.4]{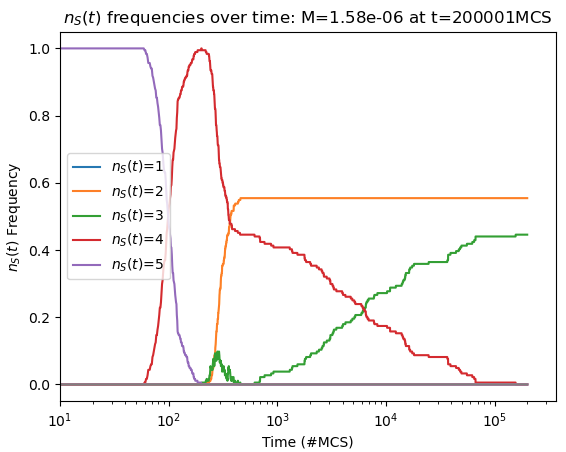}
\includegraphics[trim=0cm 0cm 0cm 0cm, clip=true,scale=0.4]{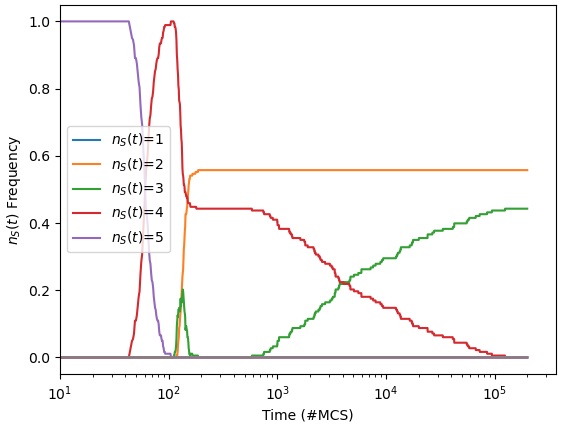}
\end{center}
\caption{Frequency distribution of species-counts to $t$$=$$200$kMCS  for RES experiments with one ablation ($N_a$$=$$1$): $L$$=$$200$; $\mu$$=$$\sigma$$=$$1.0$;
$\mu$$=$$\sigma$$=$$1.0$; 
$M$$\in$$\{9.97$$\times$$10^{-7}, 1.58$$\times$$10^{-6}, 1.58$$\times$$10^{-5}\}$ (from top to bottom). 
At each value of $M$ sampled, 175 IID experiments were performed. 
Format as for  Fig.~\ref{fig:OES_NS05_NA0_L200_A}.
}
\label{fig:RES_NS05_NA1_L200_C}
\end{figure}

\begin{figure}
\begin{center}
\includegraphics[trim=0cm 0cm 0cm 0cm, clip=true,scale=0.4]{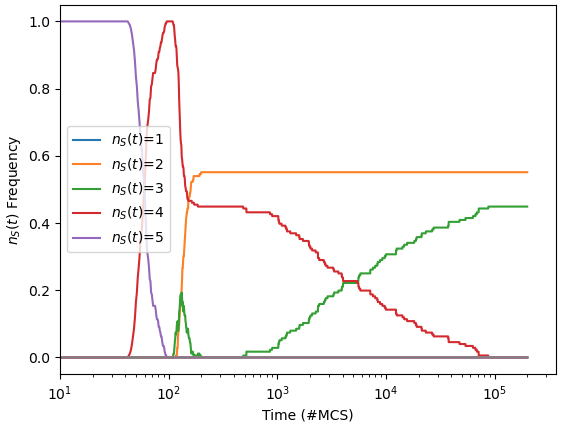}
\includegraphics[trim=0cm 0cm 0cm 0cm, clip=true,scale=0.4]{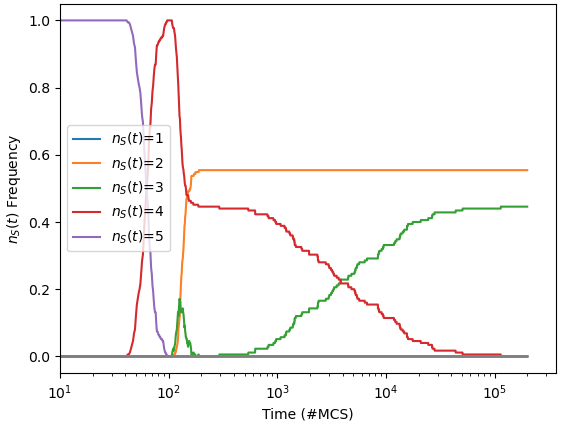}
\includegraphics[trim=0cm 0cm 0cm 0cm, clip=true,scale=0.4]{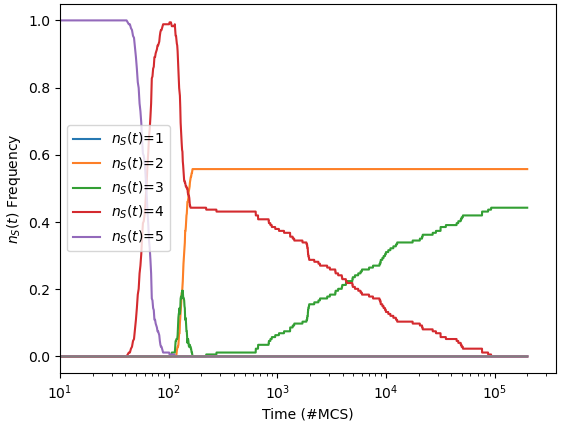}
\end{center}
\caption{Frequency distribution of species-counts to $t$$=$$200$kMCS  for RES experiments with one ablation ($N_a$$=$$1$): $L$$=$$200$; $\mu$$=$$\sigma$$=$$1.0$;
$\mu$$=$$\sigma$$=$$1.0$; 
$M$$\in$$\{1.57$$\times$$10^{-4}, 3.94$$\times$$10^{-4}, 9.89$$\times$$10^{-4}\}$ (from top to bottom). 
At each value of $M$ sampled, 175 IID experiments were performed. 
Format as for  Fig.~\ref{fig:OES_NS05_NA0_L200_A}.
}
\label{fig:RES_NS05_NA1_L200_D}
\end{figure}

Figures~\ref{fig:RES_NS05_NA1_L200_A} to~\ref{fig:RES_NS05_NA1_L200_D} show time series of the evolution the number of surviving species $n_S(t)$ over 200kMCS for the RES ESCG with one ablation (i.e., ablating either the $S_0$$\rightarrow$$S_1$ or the $S_0$$\rightarrow$$S_3$ directed edge from the dominance network, $N_a$$=$$1$), $L$$=$$200$ and $\mu$$=$$\sigma$$=$$1.0$ for various values of $M$$\in$$[ 10^{-7}, 10^{-3}]$, with 175 IID simulations executed for each value of $M$ sampled. 

Figure~\ref{fig:RES_NS05_NA1_L200_A} shows that, at low mobility ($M$$=$$10^{-7}$) the RES-based RPSLS ESCG settles to all experiments having three-species coexistence after 200kMCS: that is, $F(n_s(200\text{k})$$=$$3)=1.0$. Then, Figure~\ref{fig:RES_NS05_NA1_L200_B} shows that as $M$ is increased through the range $[2.5$$\times$$10^{-7},3.9$$\times$$10^{-7}]$, the frequency of $n_s(t)$$=$$3$ outcomes reduces, with $n_s(t)$$=$$2$ rising sharply. After that, as is seen in Figures~\ref{fig:RES_NS05_NA1_L200_C} and~\ref{fig:RES_NS05_NA1_L200_D}, successive increases in $M$ serve only to reduce the duration over which $n_s(t)$$=$$4$ is the temporarily dominant outcome; and, as these results were generated from RES with $L$$=$$200$, once $M$$>$$1.25$$\times$$10^{-5}$ there are no further changes in the system's response, because $M_{\text{max}(200)}=1/(2\times200^2)=1.25$$\times$$10^{-5}$.

\clearpage

\section*{Appendix D: Revised ES (RES), $L$$=$$200$, $N_a$$=$$0$}

As was discussed in the main text of this paper, and already illustrated in Figure~\ref{fig:RES_NS05_NA0_L150_minD}, at $L$$=$$200$ the unablated ($N_a$$=$$0$) RES-based RPSLS system shows no underflow extinctions at $t$$=$$200$kMCS for any value of $M$ sampled, and so the graphs of $n_s(t)$ vs time are not worth plotting here: each graph shows a constant flat line with $F(n_s(t)$$=$$5) =1.0$ and a zero frequency for all other outcomes over all time.

\end{document}